\documentclass[]{aa}
\usepackage{graphicx}
\usepackage{txfonts}

\newcommand{\Ms}{~M$_{\odot}$}

\newcommand{\Mstar}{$M_{\star}$}
\newcommand{\Teff}{$T_{\rm{eff}}$}

\newcommand{\Hb}{\ifmmode {\rm H}\beta \else H$\beta$\fi}

\begin{document}
       \title{Post-AGB stars as testbeds of nucleosynthesis in AGB stars}

       \author{G. Stasi\'nska\inst{1}
              \and R. Szczerba\inst{2}
              \and M. Schmidt\inst{2}
              \and N. Si\'odmiak\inst{2}
              }

       \offprints{G. Stasi\'nska, \\ \email{grazyna.stasinska@obspm.fr}}

       \institute{LUTH, Observatoire de Meudon, 5 Place Jules Janssen,
                 F-92195 Meudon Cedex, France
             \and
             N. Copernicus Astronomical Center, Rabia\'{n}ska 8,
                 87-100 Toru\'{n}, Poland
                 }

       \date{Received / Accepted }

       \abstract{We construct a data base of  125 post-AGB objects (including R CrB and extreme helium stars) with published photospheric parameters (effective temperature and gravity) and chemical composition. We estimate the masses of the post-AGB stars by comparing their position in the (log \Teff, log $g$) plane  with theoretical evolutionary tracks of different masses. We construct various diagrams, with the aim of finding clues to AGB nucleosynthesis. This is the first time that a large sample of post-AGB stars has been used in a systematic way for such a purpose and we argue that, in several respects, post-AGB stars should be more powerful than planetary nebulae to test AGB nucleosynthesis.  Our main findings are that:    the vast majority of objects which do not show evidence of N production from primary C   have  a low stellar mass ($M_{\star}$ $<$ 0.56 M$_{\odot}$);  there is no evidence that objects which did not experience 3rd dredge-up have a different stellar mass distribution than objects that did; there is clear evidence that 3rd dredge-up is more efficient at low metallicity. The sample of known post-AGB stars is likely to increase significantly in the near future thanks to the   ASTRO-F and follow-up observations, making these objects even more promising as testbeds for AGB nucleosynthesis. 
	     }

	\titlerunning{Post-AGB stars as testbeds of nucleosynthesis in ABG stars}
	\authorrunning{Stasi\'nska et al.}

       \maketitle  

%
%________________________________________________________________

\section{Introduction}

The asymptotic giant branch (AGB) phase, a late stage in the evolution of low 
and intermediate mass stars, is quite complex. During this phase, various 
nuclear processes are at work in different zones of the star, and a variety of 
mixing mechanisms take place (see e.g. Charbonnel\,\cite{Charbonnel_2002} or Lattanzio\,\cite{Lattanzio_2002} for short 
reviews). Understanding this phase is important, since the elements manufactured during 
the AGB contribute significantly to the chemical composition of galaxies.

The first models to give some predictions on the stellar yields from AGB stars 
are those by Iben \& Truran\,(\cite{Iben_and_Truran_1978}) and
Renzini \& Voli\,(\cite{Renzini_and_Voli_1981}).
Unfortunately, even nowadays, the state of stellar physics does not allow one 
to construct models entirely from first principles, and some quantities have to be set as 
free parameters. The next generation of AGB models (Groenewegen \& de 
Jong\,\cite{Groenewegen_and_de_Jong_1993}, Marigo et al.\,\cite{Marigo_etal_1996})
adjusted those free parameters (essentially mass loss rate and mixing length) 
to reproduce a few observational constraints on stellar populations. Further models 
have been constructed since then (e.g. by Forestini \& 
Charbonnel\,\cite{Forestini_and_Charbonnel_1997}, 
Boothroyd \& Sackmann\,\cite{Boothroyd_and_Sackmann_1999}, Marigo\,\cite{Marigo_2001},
 Izzard et al.\,\cite{Izzard_etal_2004}). 
All those models are so-called synthetic 
models in that, for the thermal pulse phase, they use analytical expressions 
to extrapolate certain quantities obtained from full evolutionary calculations 
up to the planetary nebula ejection. With the availability of fast computers, 
it is now possible to compute complete AGB models (Karakas et al.\,\cite{Karakas_etal_2002}, Herwig\,\cite{Herwig_2004})
that follow all the pulses in detail. Testing model predictions before using them in chemical 
evolution models of galaxies is vital, especially 
because even full evolutionary calculations are computationally difficult and still imply some  ad hoc parameters 
(mixing length, mass loss rates etc.). Indeed, as shown recently by Ventura \& D'Antona (2005a, 2005b), the predicted yields of intermediate mass stars depend strongly on the treatment adopted for convection and mass loss and on the nuclear reaction cross-sections.

As mentioned before, synthetic models published  since 1993 reproduce some 
observables (e.g. the luminosity functions of carbon-stars and of lithium-rich stars) by construction. However, additional tests are needed and are crucial to constrain AGB models. The analysis of  the chemical composition of planetary nebulae can provide some tests. Such an approach has 
been adopted recently by Marigo et al.\,(\cite{Marigo_etal_2003}) using a data base 
of 10 planetary nebulae (PNe) observed in a wide spectral range, from the far 
infrared to the ultraviolet. Previously, the chemical composition of planetary 
nebulae samples had been used in a more empirical way to test AGB 
nucleosynthesis and get an insight into the relation between planetary nebulae, 
the final products of  low and intermediate mass stars and their progenitors. 
For example, Peimbert\,(\cite{Peimbert_1978}) defined 
a category of planetary nebulae, called Type I PNe, as objects having He/H $>$ 
0.125 and N/O $>$ 0.5. Those PNe were interpreted as being born from stars 
that experienced the second dredge-up.  While the designation "Type I PNe" 
stayed in the literature, its definition changed several times (see 
Stasi\'nska\,\cite{Stasinska_2004}). 
Kingsburgh \& Barlow\,(\cite{Kingsburgh_and_Barlow_1994}) 
compared the nebular N/H ratio to the solar (C+N)/H ratio to find out which 
planetary nebulae exhibit N produced from primary C.  They also 
identified planetary nebulae whose progenitors had experienced 3rd dredge-up, 
bringing to the surface carbon produced by the triple-$\alpha$ reaction. To this
end, they constructed a (C+N+O)/H vs C/H diagram. The N/O vs O/H diagram has 
been interpreted (e.g. Henry\,\cite{Henry_1990}) as indicating that in 
some objects N is produced at the expense of O (during ON cycling). However, this
diagram, depending on authors and samples, does not always lead to such a 
straightforward interpretation 
(Kingsburgh \& Barlow\,\cite{Kingsburgh_and_Barlow_1994},
 Leisy \& Dennefeld\,\cite{Leisy_and_Dennefeld_1996}). Henry et 
al.\,(\cite{Henry_etal_2000}) have plotted PN abundances of C and N 
as a function of progenitor mass (estimated from the stellar 
remnant masses by G\'orny et al.\,\cite{Gorny_etal_1997} and Stasi\'nska et 
al.\,\cite{Stasinska_etal_1997} and converted to progenitor 
masses using an initial-final mass relation), and compared this with the 
predictions from synthetic AGB models. This was the first attempt to compare 
PNe abundances directly with the results of AGB model computations for a given 
initial mass and it was not very conclusive.

There is another category of objects that may serve for tests of AGB models. 
These are post-AGB stars, i.e. stars that have already ejected their envelope 
but are not yet hot enough to ionize it and produce a planetary nebula.  Such 
stars have several advantages with respect to PNe and constitute an excellent 
complement for testing AGB nucleosynthesis. The main advantage is that the 
stellar mass can be obtained directly from the observed stellar spectrum by 
fitting a model atmosphere that allows one to derive the stellar effective 
temperature, \Teff, and gravity $g$ (this can also be done for PNe nuclei, but  as the stars hotter and buried in the ionized gas, this is more 
difficult). The abundances of quite a variety of elements (He, Li, C, N, O, Na, 
Mg, Al, Si, S, Ca, Sc, Ti, Fe, Ni, Zn ...) can be obtained from a stellar 
atmosphere analysis. For carbon, a particularly important element in AGB evolution, the uncertainty in the estimated abundance is of the same order as for 
the other elements. This is not the case in planetary nebulae, where no strong 
carbon line exists in the optical, implying that the carbon abundance 
determination is subject to a higher uncertainty than  the oxygen or the 
nitrogen abundance determination.   Finally, post-AGB stars are expected to 
extend to smaller masses than the nuclei of planetary nebulae. Indeed, planetary 
nebulae with central star masses lower than 0.55\Ms\ do not exist since the 
nebular gas has dissipated in the interstellar medium long before the star has 
become hot enough to ionize it. On the other hand, there is no such limitation for post-AGB stars.

This paper presents the first attempt to use a large sample of post-AGB stars 
to test AGB nucleosynthesis. In Sect. 2, we describe the constitution of our 
sample. In Sect. 3 we show some empirical diagrams and propose simple 
interpretations. In Sect. 4 we summarize the main findings and outline some 
prospects.

\section{The optical sample of post-AGB stars. }

\subsection{General presentation}

As mentioned above, the post-AGB phase starts when the star has expelled its 
envelope and left the AGB branch, and is then moving to the left in the H-R diagram. 
 The end of AGB is characterized by strong mass loss, which (sometimes) can reach values of $10^{-4}$ \Ms yr$^{-1}$. For stars that have experienced intense mass loss on the AGB,  due to the large dust opacity in the envelope, the post-AGB star is first seen 
through its infrared emission due to reprocessing of the stellar radiation by 
the circumstellar dust grains. It is only when the envelope has sufficiently 
expanded and become optically thin that such stars start being optically 
visible. In this paper, we are interested  in optically visible  post-AGB 
stars.  We used the present version (Szczerba et al., in preparation) of 
our catalogue of post-AGB candidates (Szczerba et al.\,\cite{Szczerba_etal_2001}),
which now contains about  330 objects. For all the sources from this catalogue we have 
searched the available literature for stellar parameters and chemical composition. 

We then estimated the stellar masses, \Mstar, by comparison with theoretical 
evolutionary paths in the (log\,$T_{\rm eff}$, log\,$g$) plane. The theoretical 
paths we used were interpolated by G\'orny et al.\,(\cite{Gorny_etal_1997})
from the post-AGB models of Sch\"onberner\,(\cite{Schoenberner_1983})
and Bl\"ocker\,(\cite{Bloecker_1995}). 
The uncertainties in log\,$T_{\rm eff}$ and log\,$g$ induce an uncertainty 
in the derived \Mstar. The main source of uncertainties is the low accuracy of the
determinations of surface gravity, especially in the case of cool
post-AGB stars ($T_{\rm eff}$ below 10,000\,K). Since the relation between
log\,$g$ and stellar mass on the post-AGB tracks is highly nonlinear,for each object we determined  the mass $M_{{\star}{\rm min}}$ corresponding to (log $T_{\rm eff}-\Delta ({\rm log} T_{\rm eff}$), log $g-\Delta ({\rm log} g$)) and the mass $M_{{\star}{\rm max}}$ corresponding to (log $T_{\rm eff}+\Delta ({\rm log} T_{\rm eff}$), log $g+\Delta (l{\rm log} g$)). Since the derived stellar mass is a decreasing function of both $T_{\rm eff}$ and log $g$, the values of $M_{{\star}{\rm min}}$ and $M_{{\star}{\rm max}}$ should define a conservative stellar mass 
interval. \footnote{The available model grid allows us to determine masses only within the range (0.55 -- 0.94\,M$_{\odot}$).}
However, one should keep in mind that many
post-AGB stars are  pulsating stars (Gautschy \& Saio\,\cite{Gautschy_and_Saio_1996}), while such pulses are not reproduced by the evolutionary models used to derive the stellar masses.  In addition, the pulsating nature of the atmosphere may introduce errors in the abundance determinations, which are done using static atmosphere models.  A further source of uncertainty in the determination of masses comes, of course, from the model tracks themselves. All this implies that the masses of post-AGB stars are quite uncertain and model-dependent. In spite of this, we believe that the method provides useful information. It can  be improved on in the future, as better spectra become available for spectroscopic analysis and  as progress in understanding the physics of post-AGB stars is made.  In particular, when complete grids with different metallicities become available, it should be possible to account for the metallicity in the derivation of the stellar masses. For the moment, as far as one can judge from several tracks computed  by Vassiliadis \& Wood (1994) at different metallicities, the effect of metallicity in mass derivation would be completely dominated by the large error on log\,$g$.

The details of our search and the above estimates of the stellar masses are 
presented in Table 1.
%\footnote{The full version of the tables 
%are only accessible in electronic form at the CDS via anonymous ftp to 
%cdsarc.u-strasbg.fr (130.79.128.5) or via 
%http://cdsweb.u-strasbg.fr/Abstract.html.}
The objects are grouped into several subtypes: RV Tau stars (29 objects), suspected RV Tau stars (6 objects from Maas et al.\,\cite{MvWLE05}), R CrB stars (19 objects), extreme helium stars (15 objects) and all the remaining post-AGB stars (56 objects).

The RV Tau stars are highly luminous variable stars characterized by
light-curves with alternate deep and shallow minima, periods between 30 and
150 days, and F, G or K spectral types (see e.g. Preston et al. 1963). They have been identified as post-AGB stars by Jura (1986), who showed that their IRAS fluxes indicate that they have just left a phase of very rapid mass loss. 

The R CrB stars (already known for more than 200 years!) are rare
H-deficient and C-rich supergiants that undergo
irregular declines of up to 8 magnitudes when dust forms in clumps along the
line of sight (see e.g Clayton 1996 for a review). The extreme H-deficiency of
the R CrB stars suggests that some mechanism removed the entire H-rich stellar
envelope. There are two major models which explain their origin: a merger
scenario (Webbink 1984, Iben et al. 1986) or the final helium shell flash
scenario (Fujimoto 1977, Renzini 1979). There is still no consensus about
which scenario is valid (none of them can explain all the observed
properties). Only the second scenario implies a post-AGB nature. We have  included these stars in  Table 1, but we do not consider them as bona-fide post-AGB stars. Extreme helium stars, which could be evolutionarily connected to R CrB stars (see e.g. Pandey et al. 2001),  are also included in our Table and discussed together with R CrB stars.

In each subtype, the objects in Table 1 are ordered by galactic coordinates {\it l} and {\it b}. For some sources, there are several entries and in Sect.\,\ref{notes} we
briefly discuss the preferred determinations - usually, they are based on  
higher quality spectroscopic material.
The  columns 
contain the following data: (1) the object number; (2) the object coordinates 
{\it l}, {\it b}; (3) the IRAS name; (4) the HD number ; (5) other name  (either the usual name or the designation in one of the following catalogues (chosen in this order: General Catalogue of Variable Stars, LS, BD, SAO, and CD catalogues) ;
(6) the effective tempearture; (7) the error attributed to $T_{\rm eff}$; 
(8) the logarithm of the surface gravity in cm\,s$^{-2}$; (9) the error 
attributed to log\,$g$; (10)  references and notes for the collected stellar 
parameters (explanations for the abbreviations used and for notes are given at
the end of the table); (11) \Mstar\ in units of solar 
mass; (12) \Mstar$_{\rm min}$ and (13) \Mstar$_{\rm max}$.

%********************************* table 1 *************************
\begin{table*}[t]{}
\caption[]{List of post-AGB objects with stellar parameters determined from model 
atmosphere techniques.}
\label{table1}
\centering
\begin{tabular}{l@{ }l@{ }l@{ }l@{  }l@{  }r@{  }r@{  }l@{  }l@{  }l@{  }r@{  }r@{  }r@{  }}
\hline
\noalign{\smallskip}
 No & l~~       b    &IRAS         & HD      & other name          & $T_{\rm eff}$& err.  & log\,$g$ & err.  & notes          &$M_{\star}$&$M_{{\star}{\rm min}}$&$M_{{\star}{\rm max}}$\\   
(1) & (2)            &(3)          & (4)     & (5)                 &(6)           &(7)    &(8)       &(9)    &(10)            & (11)    &(12)          &(13)          \\
\noalign{\smallskip}
\hline
\noalign{\smallskip}
  1 & ~006.72 $-$10.37 &~18384$-$2800 & ~172481  & ~V4728 Sgr           & 7250         & 200   & ~1.5      & 0.25  & AFGM01         &  $<$0.550  &  $<$0.550  &  $<$0.550        \\
    &                 &              &         &                     & 7250         & 250   & ~1.5      & 0.5   & RvW01          &  $<$0.550  &  $<$0.550  &     0.554        \\
  2 & ~007.96 $+$26.71 &~~F16277$-$0724 & ~148743  & ~BD$-$07 4305        & 7200         & 500   & ~0.5      & 0.3   & LBL90$^{1)}$   &     0.828  &     0.627  &  $>$0.940        \\
  3 & ~013.23 $+$12.17 &~17279$-$1119 & ~158616  & ~V340 Ser            & 7550         & 150   & ~0.75     & 0.25  & vW97$^{2)}$    &     0.632  &     0.576  &     0.874        \\
    &                 &              &         &                     & 7300         & 200   & ~1.5      & 0.25~~& AFGM01         &  $<$0.550  &  $<$0.550  &  $<$0.550        \\
  4 & ~016.45 $-$50.43 &              &         & ~BPS CS 29493-0046   &20000         &~~3000 & ~3.0      & 0.3   & KDK97          &     0.550  &  $<$0.550  &     0.550        \\
  5 & ~023.98 $-$21.04 &~19500$-$1709 & ~187885  & ~V5112 Sgr           & 8000         & 250   & ~1.0      & 0.5   & vWR00          &     0.599  &  $<$0.550  &     0.933        \\
    &                 &              &         &                     & 7850         & 150   & ~0.75     & 0.25  & vWWW96         &     0.673  &     0.599  &     0.926        \\
  6 & ~029.18 $-$21.26 &~19590$-$1249 &         & ~LS IV $-$12 111     &20500         & 500   & ~2.35     & 0.2   & RDM03$^{3)}$   &     0.757  &     0.625  &     0.931        \\
    &                 &              &         &                     &23750         &1000   & ~2.7      & 0.2   & MCD92          &     0.649  &     0.612  &     0.826        \\
  7 & ~030.60 $-$21.53 &~20023$-$1144 & ~190390  & ~V1401 Aql           & 6600         & 500   & ~1.6      & 0.3   & LBL90$^{1)}$   &  $<$0.550  &  $<$0.550  &  $<$0.550        \\
  8 & ~031.33 $-$43.48 &              &         & ~PHL 1580            &24000         &1000   & ~3.6      & 0.2   & CDK91          &  $<$0.550  &  $<$0.550  &  $<$0.550        \\
    &                 &              &         &                     &21500         &1500   & ~3.0      & 0.35  & KL86           &     0.552  &  $<$0.550  &     0.578        \\
  9 & ~033.16 $-$48.12 &              &         & ~PHL 174             &18000         &1000   & ~2.7      & 0.2   & CDK91          &     0.552  &     0.550  &     0.556        \\
 10 & ~035.62 $-$04.96 &~19114$+$0002 & ~179821  & ~V1427 Aql           & 6750         & 150   & ~0.5      & 0.5   & RH99           &     0.660  &     0.553  &  $>$0.940        \\
    &                 &              &         &                     & 6800         & 250   & ~1.3      & 0.5   & ZKP96          &  $<$0.550  &  $<$0.550  &     0.559        \\
    &                 &              &         &                     & 5660         & 100   &$-$1.0      & 0.5   & TPJ00          &  $>$0.940  &  $>$0.940  &  $>$0.940        \\
 11 & ~040.51 $-$10.09 &~19386$+$0155 &         & ~V1648 Aql           & 6800         & 100   & ~1.4      & 0.2   & PLM04          &  $<$0.550  &  $<$0.550  &  $<$0.550        \\
 12 & ~043.06 $+$32.36 &              &         & ~PG 1704+222         &20500         &1000   & ~3.0      & 0.2   & CTM93          &     0.551  &  $<$0.550  &     0.554        \\
    &                 &              &         &                     &17600         & 400   & ~2.7      & 0.1   & MH98$^{4)}$    &     0.551  &     0.550  &     0.553        \\
 13 & ~043.23 $-$57.13 &~22327$-$1731 & ~213985  & ~HM Aqr              & 8200         &       & ~1.5      &       & vW95           &  $<$0.550  &            &                  \\
 14 & ~050.67 $+$19.79 &~18062$+$2410 & ~341617  & ~V886 Her            &23000         &1000   & ~2.6      & 0.2   & MRK02          &     0.692  &     0.621  &     0.868        \\
    &                 &              &         &                     &23000         & 200   & ~3.0      & 0.25  & AFGM01         &     0.558  &     0.550  &     0.614        \\
    &                 &              &         &                     &22000         & 200   & ~3.0      & 0.5   & PGS00          &     0.554  &  $<$0.550  &     0.697        \\
    &                 &              &         &                     &20750         & 500   & ~2.35     & 0.2   & RDM03$^{3)}$   &     0.813  &     0.631  &  $>$0.940        \\
 15 & ~051.43 $+$23.19 &~17534$+$2603 & ~163506  & ~89 Her              & 6550         & 500   & ~0.6      & 0.3   & LBL90$^{1)}$   &     0.611  &     0.561  &     0.670        \\
 16 & ~052.73 $+$50.79 &              &         & ~BD$+$33 2642        &20200         & 500   & ~2.9      & 0.1   & NHK94          &     0.552  &     0.551  &     0.554        \\
 17 & ~053.84 $+$20.18 &~18095$+$2704 &         & ~V887 Her            & 6600         & 300   & ~1.05     & 0.5   & K95$^{5)}$     &     0.550  &  $<$0.550  &     0.607        \\
 18 & ~066.18 $+$18.58 &              & ~172324  & ~V534 Lyr            &11250         & 200   & ~2.5      & 0.25  & AFGM01$^{6)}$  &  $<$0.550  &  $<$0.550  &  $<$0.550        \\
 19 & ~067.16 $+$02.73 &~19475$+$3119 & ~331319  & ~LS II $+$31 9       & 7250         & 100   & ~0.5      & 0.3   & KPT02          &     0.841  &     0.611  &  $>$0.940        \\
    &                 &              &         &                     & 7750         & 200   & ~1.0      & 0.25  & AFGM01         &     0.578  &     0.553  &     0.632        \\
 20 & ~077.13 $+$30.87 &~17436$+$5003 & ~161796  & ~V814 Her            & 6600         & 500   & ~0.25     & 0.3   & LBL90$^{1)}$   &     0.915  &     0.698  &  $>$0.940        \\
    &                 &              &         &                     & 7100         & 100   & ~0.5      & 0.3   & KPT02          &     0.785  &     0.604  &  $>$0.940        \\
 21 & ~080.17 $-$06.50 &              &         & ~Egg Nebula          & 6500         & 200   & ~0.0      & 0.3   & KSP00b         &  $>$0.940  &     0.897  &  $>$0.940        \\
 22 & ~096.75 $-$11.56 &~22223$+$4327 &         & ~BD$+$42 4388        & 6500         & 250   & ~1.0      & 0.5   & vWR00          &     0.551  &  $<$0.550  &     0.614        \\
    &                 &              &         &                     & 6500         & 350   & ~1.0      & 0.5   & DvWW98         &     0.551  &  $<$0.550  &     0.609        \\
 23 & ~098.41 $-$16.73 &              &         & ~BD$+$39 4926        & 7500         &       & ~0.5      &       & K73            &     0.894  &            &                  \\
 24 & ~103.35 $-$02.52 &~22272$+$5435 & ~235858  & ~V354 Lac            & 5750         & 150   & ~0.5      & 0.5   & RLG02          &     0.574  &  $<$0.550  &     0.888        \\
    &                 &              &         &                     & 5600         & 250   & ~0.5      & 0.5   & ZKP95          &     0.561  &  $<$0.550  &     0.805        \\
 25 & ~113.86 $+$00.59 &~23304$+$6147 &         &                     & 6750         & 250   & ~0.5      & 0.5   & vWR00          &     0.660  &     0.554  &  $>$0.940        \\
    &                 &              &         &                     & 5900         & 200   & ~0.0      & 0.5   & KSP00a         &  $>$0.940  &     0.606  &  $>$0.940        \\
 26 & ~123.57 $+$16.59 &~01005$+$7910 &         &                     &21000         & 500   & ~3.0      & 0.3   & KYM02          &     0.551  &  $<$0.550  &     0.576        \\
 27 & ~133.73 $+$01.50 &~Z02229$+$6208&         &                     & 5500         & 250   & ~0.5      & 0.25  & RBH99          &     0.558  &     0.551  &     0.604        \\
 28 & ~161.98 $+$19.59 &~06338$+$5333 & ~46703   & ~V382 Aur            & 6000         & 150   & ~0.4      & 0.3   & LB84           &     0.620  &     0.558  &     0.867        \\
 29 & ~166.24 $-$09.05 &~04296$+$3429 &         &                     & 7000         & 250   & ~1.0      & 0.5   & vWR00          &     0.554  &  $<$0.550  &     0.660        \\
    &                 &              &         &                     & 6300         & 250   & ~0.0      & 0.2   & KSP99          &  $>$0.940  &     0.938  &  $>$0.940        \\
    &                 &              &         &                     & 7000         & 350   & ~1.0      & 0.3   & DvWW98         &     0.554  &  $<$0.550  &     0.591        \\
 30 & ~172.95 $-$05.50 &              &         & ~Barnard 29          &20000         &1000   & ~3.0      & 0.1   & CDK94          &     0.550  &     0.550  &     0.550        \\
 31 & ~173.86 $-$82.41 &              &         & ~BPS CS 22946$-$0005 &20000         &3000   & ~2.7      & 0.3   & KDK97          &     0.565  &     0.558  &     0.571        \\
 32 & ~188.86 $-$14.29 &~05113$+$1347 &         &                     & 5250         & 150   & ~0.25     & 0.5   & RLG02          &     0.604  &     0.550  &  $>$0.940        \\
 33 & ~196.19 $-$12.14 &~05341$+$0852 &         &                     & 6500         & 250   & ~1.0      & 0.5   & vWR00          &     0.551  &  $<$0.550  &     0.614        \\
 34 & ~204.67 $+$07.57 &~07008$+$1050 & ~52961   & ~PS Gem              & 6000         & 500   & ~0.5      & 0.5   & WvWB91         &     0.599  &     0.551  &     0.860        \\
 35 & ~206.75 $+$09.99 &~07134$+$1005 & ~56126   & ~LS VI $+$10 15      & 7250         & 250   & ~0.5      & 0.5   & HR03           &     0.841  &     0.563  &  $>$0.940        \\
    &                 &              &         &                     & 7250         & 250   & ~0.5      & 0.5   & vWR00          &     0.841  &     0.563  &  $>$0.940        \\
    &                 &              &         &                     & 7000         & 300   & ~0.1      & 0.5   & K95            &  $>$0.940  &     0.697  &  $>$0.940        \\
 36 & ~208.93 $+$17.07 &~07430$+$1115 &         &                     & 6000         & 250   & ~1.0      & 0.25  & RBH99          &  $<$0.550  &     0.540  &     0.551        \\
 37 & ~215.44 $-$00.13 &~06530$-$0213 &         &                     & 6900         & 250   & ~1.0      & 0.5   & HR03           &     0.553  &     0.540  &     0.643        \\
    &                 &              &         &                     & 7250         & 250   & ~1.0      & 0.5   & RvWG04         &     0.557  &  $<$0.550  &     0.722        \\
 38 & ~218.97 $-$11.76 &~06176$-$1036 & ~44179   & ~Red Rectangle       & 7500         &       & ~0.8      &       & WvWTW92~~      &     0.617  &            &                  \\
  \noalign{\smallskip}
  \hline
\end{tabular}
\end{table*}
\begin{table*}[t]{}
	\centerline{Table\,\ref{table1}. (continued)}
\centering
\begin{tabular}{l@{ }l@{ }l@{ }l@{  }l@{  }r@{  }r@{  }l@{  }l@{  }l@{  }r@{  }r@{  }r@{  }}
\hline
\noalign{\smallskip}
 No & l~~       b    &IRAS         & HD      & other name          & $T_{\rm eff}$& err.  & log\,$g$ & err.  & notes          &$M_{\star}$&$M_{{\star}{\rm min}}$&$M_{{\star}{\rm max}}$\\   
(1) & (2)            &(3)          & (4)     & (5)                 &(6)           &(7)    &(8)       &(9)    &(10)            & (11)    &(12)          &(13)          \\
\noalign{\smallskip}
\hline
\noalign{\smallskip}
 39 & ~236.57 $-$05.37 &~07140$-$2321 &         & ~SAO 173329          & 6750         & 250   & ~1.25     & 0.25  & vW97$^{2)}$    &  $<$0.550  &  $<$0.550  &     0.551        \\
 40 & ~260.83 $-$05.07 &~08143$-$4406 &         &                     & 7150         & 100   & ~1.35     & 0.15  & RvWG04$^{7)}$  &  $<$0.550  &  $<$0.550  &     0.550        \\
 41 & ~264.55 $+$72.47 &              & ~105262  & ~BD$+$13 2491        & 9250         & 250   & ~1.8      & 0.25  & RPS96          &  $<$0.550  &  $<$0.550  &     0.551        \\
 42 & ~266.85 $+$22.93 &~10158$-$2844 & ~ 89353  & ~AG Ant              & 7600         & 400   & ~1.05     & 0.1   & WvWB91         &     0.559  &     0.558  &     0.559        \\
 43 & ~269.97 $-$34.08 &              &         & ~CPD-61 455          &25000         &1000   & ~3.6      & 0.2   & HDK96          &  $<$0.550  &  $<$0.550  &  $<$0.550        \\
 44 & ~290.54 $-$01.95 &~11000$-$6153 & ~ 95767  & ~LS 2105             & 7300         & 300   & ~1.5      & 0.25  & vW97$^{2)}$    &  $<$0.550  &  $<$0.550  &  $<$0.550        \\
 45 & ~293.03 $+$05.94 &~11385$-$5517 & ~101584  & ~V885 Cen            & 8500         & 500   & ~1.5      & 0.5   & SPG99          &     0.550  &  $<$0.550  &     0.599        \\
 46 & ~295.48 $+$29.87 &              & ~107369  & ~SAO 203367          & 7600         & 400   & ~1.5      & 0.25  & vW97$^{2)}$    &  $<$0.550  &  $<$0.550  &     0.550        \\
 47 & ~298.25 $+$15.48 &~12222$-$4652 & ~108015  & ~SAO 223420          & 6800         & 200   & ~1.25     & 0.25  & vW97$^{2)}$    &  $<$0.550  &  $<$0.550  &     0.551        \\
 48 & ~298.30 $+$08.67 &~12175$-$5338 &         & ~V1024 Cen           & 7350         & 150   & ~0.75     & 0.25  & vW97$^{2)}$    &     0.620  &     0.563  &     0.828        \\
 49 & ~299.02 $-$43.77 &              &         & ~LB 3219             &21250         &1000   & ~2.8      & 0.2   & MCD92          &     0.567  &     0.555  &     0.603        \\
 50 & ~304.34 $+$36.40 &~12538$-$2611 & ~112374  & ~LN Hya              & 6000         & 200   & ~1.0      & 0.25  & GAFP97         &  $<$0.550  &  $<$0.550  &     0.552        \\
    &                 &              &         &                     & 6000         & 275   & ~0.6      & 0.3   & LLB83          &     0.568  &     0.551  &     0.624        \\
 51 & ~309.07 $+$15.18 &              & ~116745  & ~Fehrenbach's star   & 6950         &  75   & ~1.15     & 0.1   & GW92           &     0.550  &  $<$0.550  &     0.552        \\
 52 & ~317.11 $+$53.11 &              &         & ~BPS CS 22877-0023~~ &20000         &3000   & ~3.0      & 0.3   & KDK97          &     0.550  &  $<$0.550  &     0.550        \\
    &                 &              &         &                     &16400         & 700   & ~2.5      & 0.15  & MH98$^{4)}$    &     0.553  &     0.551  &     0.557        \\
 53 & ~325.04 $+$08.65 &~15039$-$4806 & ~133656  & ~LS 3309             & 8000         & 200   & ~1.25     & 0.25~~& vWOT96$^{8)}$  &     0.554  &  $<$0.550  &     0.582        \\
 54 & ~326.77 $-$07.49 &~16206$-$5956 &         & ~LS 3591             &11200         &1000   & ~2.3      & 1.0   & GP03           &  $<$0.550  &  $<$0.550  &     0.624        \\
    &                 &              &         &                     & 8500         & 200   & ~0.9      & 0.2   & VSL98          &     0.660  &     0.609  &     0.874        \\
 55 & ~330.64 $-$03.67 &              &         & ~LS 3593             & 9300         & 200   & ~1.7      & 0.2   & VSL98          &  $<$0.550  &  $<$0.550  &     0.553        \\
 56 & ~345.58 $-$07.30 &              &         & ~[DSH2001] 279-19    &24000         &1000   & ~3.3      & 0.2   & MDS04          &     0.550  &  $<$0.550  &     0.553        \\
\noalign{\smallskip}
\multicolumn{13}{l}{\textbf{RV Tau stars}}
\\
\noalign{\smallskip}
 57 & ~020.72 $+$17.57 &              &         & ~V453 Oph            & 5800         & 200   & ~0.75     & 0.25  & GLG98          &     0.552  &  $<$0.550  &     0.561        \\
    &                 &              &         &                     & 6250         &       & ~1.5      &       & RDvW04         &  $<$0.550  &            &                  \\
 58 & ~021.48 $+$28.63 &              &         & ~TT Oph              & 4800         & 200   & ~0.5      & 0.25  & GLG00          &     0.550  &  $<$0.550  &     0.555        \\
 59 & ~026.52 $-$05.42 &~18564$-$0814 &         & ~AD Aql              & 6300         & 150   & ~1.25     & 0.15  & GLG98          &  $<$0.550  &  $<$0.550  &  $<$0.550        \\
 60 & ~044.10 $-$61.56 &              & ~216457  & ~DS Aqr              & 6500         & 200   & ~1.0      & 0.25  & GLG00          &     0.551  &  $<$0.550  &     0.558        \\
    &                 &              &         &                     & 6500         & 200   & ~2.0      & 0.25  & GLG98          &  $<$0.550  &  $<$0.550  &  $<$0.550        \\
    &                 &              &         &                     & 5750         &       & ~0.5      &       & RDvW04         &     0.574  &            &                  \\
 61 & ~050.49 $+$14.24 &~18281$+$2149 & ~170756  & ~AC Her              & 5900         & 150   & ~1.13     & 0.15  & GLG98$^{6)}$   &  $<$0.550  &  $<$0.550  &  $<$0.550        \\
 62 & ~057.53 $-$09.75 &~20117$+$1634 & ~192388  & ~R Sge               & 5000         & 200   & $-$0.25     & 0.2   & GLG97a $^{9)}$ &     0.926  &     0.806  &  $>$0.940        \\
 63 & ~058.44 $-$07.46 &~20056$+$1834 &         & ~QY Sge              & 5850         & 200   & ~0.7      & 0.25  & RGL02          &     0.554  &  $<$0.550  &     0.579        \\
 64 & ~060.73 $+$06.94 &~19163$+$2745 &         & ~EP Lyr              & 6200         & 200   & ~1.4      & 0.2   & GLG97a$^{9)}$  &  $<$0.550  &  $<$0.550  &  $<$0.550        \\
 65 & ~076.75 $-$11.78 &              &         & ~V360 Cyg            & 5275         & 200   & ~1.38     & 0.25  & GLG98$^{6)}$   &  $<$0.550  &  $<$0.550  &  $<$0.550        \\
 66 & ~148.26 $+$05.26 &~04166$+$5719 &         & ~TW Cam              & 4800         & 200   & ~0.0      & 0.25  & GLG00          &     0.622  &     0.576  &     0.773        \\
 67 & ~174.77 $-$12.19 &~04440$+$2605 &         & ~RV Tau              & 4500         & 200   & ~0.0      & 0.25  & GLG00          &     0.599  &     0.557  &     0.639        \\
 68 & ~188.06 $+$01.30 &~06054$+$2237 & ~41870   & ~SS Gem              & 5400         & 150   & ~0.2      & 0.2   & GLG97b$^{9)}$  &     0.623  &     0.585  &     0.722        \\
 69 & ~195.41 $-$03.42 &~06034$+$1354 &         & ~DY Ori              & 5900         & 200   & ~1.5      & 0.2   & GLG97a$^{9)}$  &  $<$0.550  &  $<$0.550  &  $<$0.550        \\
 70 & ~199.39 $-$04.56 &~06072$+$0953 &         & ~CT Ori              & 5750         & 150   & ~1.0      & 0.2   & GLG97b$^{9)}$  &  $<$0.550  &  $<$0.550  &     0.550        \\
 71 & ~217.80 $+$09.95 &~07331$+$0021 &         & ~AI CMi              & 4500         &       & ~0.0      &       & KP96           &     0.599  &            &                  \\
 72 & ~226.14 $+$04.15 &~07284$-$0940 & ~59693   & ~U Mon               & 5000         & 200   & ~0.0      & 0.25  & GLG00          &     0.645  &     0.599  &     0.873        \\
 73 & ~253.02 $-$03.00 &~08011$-$3627 &         & ~AR Pup              & 6300         & 200   & ~1.5      & 0.2   & GLG97a$^{9)}$  &  $<$0.550  &  $<$0.550  &  $<$0.550        \\
 74 & ~282.42 $-$09.24 &~09256$-$6324 & ~82084/5 & ~IW Car              & 6700         & 200   & ~2.0      & 0.25  & GRL94          &  $<$0.550  &  $<$0.550  &  $<$0.550        \\
 75 & ~293.17 $-$17.24 &~09538$-$7622 &         &                     & 5500         & 250   & ~1.0      & 0.5   & MvWLE05        &  $<$0.550  &  $<$0.550  &     0.554        \\
 76 & ~295.25 $+$16.82 &~12067$-$4508 & ~105578  & ~RU Cen              & 6000         & 250   & ~1.0      & 0.50  & MvWW02         &  $<$0.550  &  $<$0.550  &     0.574        \\
 77 & ~297.87 $+$13.36 &~12185$-$4856 & ~107439  & ~SX Cen              & 6250         & 250   & ~1.5      & 0.50  & MvWW02         &  $<$0.550  &  $<$0.550  &  $<$0.550        \\
 78 & ~313.90 $-$08.68 &~14524$-$6838 & ~131356  & ~EN TrA              & 6150         &  75   & ~1.25     & 0.25  & vW97$^{2)}$    &  $<$0.550  &  $<$0.550  &  $<$0.550        \\
 79 & ~330.84 $+$57.77 &~13467$-$0141 & ~120408  & ~CE Vir              & 4300         & 100   & ~0.3      & 0.2   & GLG97b$^{9)}$  &     0.550  &  $<$0.550  &     0.555        \\
 80 & ~331.86 $-$13.78 &~17250$-$5951 &         & ~UY Ara              & 5500         & 200   & ~0.3      & 0.25  & GLG00          &     0.610  &     0.559  &     0.693        \\
 81 & ~339.79 $-$04.68 &~17038$-$4815 &         &                     & 4750         & 250   & ~0.5      & 0.5   & MvWLE05        &     0.550  &  $<$0.550  &     0.599        \\
 82 & ~344.12 $+$26.45 &              &         & ~BT Lib              & 5800         & 200   & ~1.4      & 0.25  & GLG00$^{10)}$  &  $<$0.550  &  $<$0.550  &  $<$0.550        \\
 83 & ~345.49 $-$04.99 &~17243$-$4348 &         & ~LR Sco              & 6750         & 250   & ~0.25     & 0.5   & MvWLE05$^{4)}$ &     0.939  &     0.604  &  $>$0.940        \\
 84 & ~345.54 $+$10.26 &~16230$-$3410 &         &                     & 6250         & 250   & ~1.0      & 0.5   & MvWLE05        &     0.550  &  $<$0.550  &     0.599        \\
 85 & ~345.65 $-$04.69 &~17233$-$4330 &         &                     & 6250         & 250   & ~1.5      & 0.5   & MvWLE05        &  $<$0.550  &  $<$0.550  &  $<$0.550        \\
  \noalign{\smallskip}
  \hline
\end{tabular}
\end{table*}
\begin{table*}[t]{}
	\centerline{Table\,\ref{table1}. (continued)}
\centering
\begin{tabular}{l@{ }l@{ }l@{ }l@{  }l@{  }r@{  }r@{  }l@{  }l@{  }l@{  }r@{  }r@{  }r@{  }}
\hline
\noalign{\smallskip}
 No & l~~       b    &IRAS         & HD      & other name          & $T_{\rm eff}$& err.  & log\,$g$ & err.  & notes          &$M_{\star}$&$M_{{\star}{\rm min}}$&$M_{{\star}{\rm max}}$\\   
(1) & (2)            &(3)          & (4)     & (5)                 &(6)           &(7)    &(8)       &(9)    &(10)            & (11)    &(12)          &(13)          \\
\noalign{\smallskip}
\hline
\noalign{\smallskip}
\multicolumn{13}{l}{\textbf{suspected RV Tau stars from Maas et al.\,\cite{MvWLE05}}}
\\
\noalign{\smallskip}
 86 & ~033.59 $-$07.22 &~19157$-$0247 &         & ~LS IV $-$02 29      & 7750         & 250   & ~1.0      & 0.5   & MvWLE05        &     0.578  &  $<$0.550  &     0.894        \\
 87 & ~039.02 $-$03.49 &~19125$+$0343 &         & ~LS IV $+$03 18      & 7750         & 250   & ~1.0      & 0.5   & MvWLE05        &     0.578  &  $<$0.550  &     0.894        \\
 88 & ~254.58 $+$12.94 &~09060$-$2807 &         & ~BZ Pyx              & 6500         & 250   & ~1.5      & 0.5   & MvWLE05        &  $<$0.550  &  $<$0.550  &     0.550        \\
 89 & ~265.50 $+$00.39 &~08544$-$4431 &         &                     & 7250         & 250   & ~1.5      & 0.5   & MvWLE05        &  $<$0.550  &  $<$0.550  &     0.554        \\
 90 & ~271.51 $-$00.50 &~09144$-$4933 &         &                     & 5750         & 250   & ~0.5      & 0.5   & MvWLE05        &     0.574  &  $<$0.550  &     0.860        \\
 91 & ~327.82 $+$00.63 &~15469$-$5311 &         &                     & 7500         & 250   & ~1.5      & 0.5   & MvWLE05        &  $<$0.550  &  $<$0.550  &     0.557        \\
\noalign{\smallskip}
\multicolumn{13}{l}{\textbf{R CrB stars}}
\\
\noalign{\smallskip}
 92 & ~002.41 $+$07.51 &              &         & ~V2552 Oph           & 6750         & 250   & ~0.5      & 0.5   & RL03 $^{11)}$  &     0.660  &     0.554  &  $>$0.940        \\
 93 & ~002.52 $-$05.97 &~18119$-$2943 & ~317333  & ~VZ Sgr              & 7000         & 250   & ~0.5      & 0.5   & AGL00          &     0.722  &     0.557  &  $>$0.940        \\
 94 & ~004.43 $-$19.45 &~19132$-$3336 & ~180093  & ~RY Sgr              & 7250         & 250   & ~0.75     & 0.5   & AGL00          &     0.616  &     0.551  &  $>$0.940        \\
 95 & ~005.81 $-$03.78 &~18103$-$2547 &         & ~V3795 Sgr           & 8000         & 250   & ~1.0      & 0.5   & AGL00          &     0.599  &  $<$0.550  &     0.933        \\
 96 & ~008.31 $-$05.24 &~18211$-$2417 &         & ~GU Sgr              & 6250         & 250   & ~0.5      & 0.5   & AGL00          &     0.614  &     0.551  &  $>$0.940        \\
 97 & ~023.83 $-$02.92 &~18425$-$0928 &         & ~FH Sct              & 6250         & 250   & ~0.25     & 0.5   & AGL00          &     0.833  &     0.564  &  $>$0.940        \\
 98 & ~045.05 $+$50.98 &~15465$+$2818 & ~141527  & ~R CrB               & 6750         & 250   & ~0.5      & 0.5   & AGL00          &     0.660  &     0.554  &  $>$0.940        \\
    &                 &              &         &                     & 7000         & 150   & ~0.5      & 0.25  & RGA90          &     0.722  &     0.611  &  $>$0.940        \\
 99 & ~070.45 $+$02.20 &~19577$+$3351 &         & ~V482 Cyg            & 6500         & 250   & ~0.5      & 0.5   & AGL00          &     0.629  &     0.552  &  $>$0.940        \\
100 & ~109.52 $-$00.39 &~23001$+$5920 &         & ~UV Cas              & 7250         & 250   & ~0.5      & 0.5   & AGL00          &     0.841  &     0.563  &  $>$0.940        \\
101 & ~149.84 $+$01.12 &              & ~25878   & ~XX Cam              & 7250         & 250   & ~0.75     & 0.5   & AGL00          &     0.616  &     0.551  &  $>$0.940        \\
102 & ~188.86 $-$04.42 &~05461$+$1903 & ~247925  & ~SU Tau              & 6500         & 250   & ~0.5      & 0.5   & AGL00          &     0.629  &     0.552  &  $>$0.940        \\
103 & ~279.06 $+$20.12 &              &         & ~UX Ant              & 7000         & 250   & ~0.5      & 0.5   & AGL00          &     0.722  &     0.557  &  $>$0.940        \\
104 & ~301.74 $+$08.32 &~12404$-$5415 &         & ~UW Cen              & 7500         & 250   & ~1.0      & 0.5   & AGL00          &     0.563  &  $<$0.550  &     0.841        \\
105 & ~304.42 $-$02.68 &~13025$-$6514 &         & ~Y Mus               & 7250         & 250   & ~0.75     & 0.5   & AGL00          &     0.616  &     0.551  &  $>$0.940        \\
106 & ~307.96 $+$08.29 &~13224$-$5359 &         & ~DY Cen              &19500         & 500   & ~2.15     & 0.1   & JH93           &     0.902  &     0.853  &     0.938        \\
107 & ~327.22 $-$06.92 &~16200$-$5913 &         & ~RT Nor              & 7000         & 250   & ~1.5      & 0.5   & AGL00          &  $<$0.550  &  $<$0.550  &     0.552        \\
108 & ~332.44 $-$03.57 &~16287$-$5309 &         & ~RZ Nor              & 6750         & 250   & ~0.75     & 0.5   & AGL00          &     0.581  &  $<$0.550  &     0.895        \\
109 & ~347.53 $-$14.14 &~18151$-$4634 &         & ~RS Tel              & 6750         & 250   & ~1.25     & 0.5   & AGL00          &  $<$0.550  &  $<$0.550  &     0.564        \\
110 & ~357.66 $-$15.65 &~18441$-$3812 & ~173539  & ~V CrA               & 6250         & 250   & ~0.5      & 0.5   & AGL00          &     0.614  &     0.551  &  $>$0.940        \\
\noalign{\smallskip}
\multicolumn{13}{l}{\textbf{extreme helium stars}}
\\
\noalign{\smallskip}
111 & ~006.01 $+$26.02 &              &         & ~V2205 Oph           &22700         &1200   & ~2.55     & 0.1   & JH92           &     0.715  &     0.705  &     0.721        \\
112 & ~007.82 $+$05.07 &              &         & ~LS 4357             &16130         & 500   & ~2.00     & 0.25  & JHHJ98         &     0.675  &     0.605  &     0.913        \\
113 & ~020.91 $-$08.31 &              &         & ~V4732 Sgr           & 9500         & 250   & ~0.9      & 0.2   & PRL01          &     0.902  &     0.699  &  $>$0.940        \\
    &                 &              &         &                     & 9000         & 250   & ~1.0      & 0.5   & AGL00          &     0.659  &     0.554  &  $>$0.940        \\
114 & ~024.41 $+$12.50 &              &         & ~V2244 Oph           &12750         & 250   & ~1.75     & 0.25  & PRL01          &     0.612  &     0.558  &     0.732        \\
115 & ~026.55 $+$10.09 &              &         & ~No Ser              &11750         & 250   & ~2.30     & 0.4   & PRL01          &  $<$0.550  &  $<$0.550  &     0.553        \\
116 & ~031.33 $+$33.28 &              &         & ~V652 Her            &24550         & 500   & ~3.68     & 0.05  & JHH99          &  $<$0.550  &  $<$0.550  &  $<$0.550        \\
117 & ~049.87 $-$25.21 &              &         & ~FQ Aqr              & 8750         & 250   & ~0.75     & 0.25  & PRL01          &     0.908  &     0.659  &  $>$0.940        \\
    &        $ $      &              &         &                     & 8500         & 250   & ~1.5      & 0.5   & AGL00          &     0.550  &  $<$0.550  &     0.611        \\
118 & ~068.90 $+$04.76 &              & ~225642  & ~V1920 Cyg           &16180         & 500   & ~2.0      & 0.25  & PLRJ04         &     0.681  &     0.606  &     0.917        \\
119 & ~222.95 $-$04.18 &              &         & ~LS 99               &15330         & 500   & ~1.90     & 0.25  & JHHJ98         &     0.688  &     0.607  &     0.919        \\
120 & ~235.21 $+$54.44 &              &         & ~DN Leo              &16800         & 600   & ~2.55     & 0.2   & Heb83          &     0.553  &     0.550  &     0.560        \\
121 & ~309.95 $-$04.25 &              &         & ~BX Cir              &23300         & 700   & ~3.35     & 0.10  & DJH98          &  $<$0.550  &  $<$0.550  &  $<$0.550        \\
122 & ~317.65 $+$14.18 &              & ~124448  & ~V821 Cen            &15500         & 800   & ~2.1      & 0.2   & PLRJ04         &     0.610  &     0.576  &     0.634        \\
123 & ~338.13 $-$18.71 &              & ~168476  & ~PV Tel              &14000         & 500   & ~1.5      & 0.2   & WS81           &  $>$0.940  &     0.845  &  $>$0.940        \\
124 & ~344.19 $-$08.84 &              &         & ~CD-46 11775         &18000         & 700   & ~2.00     & 0.1   & Jef93          &     0.910  &     0.883  &     0.930        \\
125 & ~348.17 $+$17.78 &              & ~144941  &                     &23200         & 500   & ~3.90     & 0.2   & HJ97           &  $<$0.550  &  $<$0.550  &  $<$0.550        \\
	\noalign{\smallskip}
	\hline
\end{tabular}
\begin{list}{}{}
\item[Notes in column (8):]
\item[~~]
$^{1)}$ -- average from Table\,4 of Luck et al.\,(\cite{LBL90});   
$^{2)}$ -- average from two models of Van Winckel\,\cite{vW97};
$^{3)}$ -- an error of 500\,K for $T_{\rm eff}$ has been assumed;
$^{4)}$ -- average of models for different time of observations;
$^{5)}$ -- average of models for two spectra; 
$^{6)}$ -- average from two phases;
$^{7)}$ -- average from two models of Reyniers et al.\,(\cite{RvWG04});
$^{8)}$ -- average from two models of Van Winckel et al.\,(\cite{vWOT96});
$^{9)}$ -- avearage value;
$^{10)}$ -- log\,($g$) is averaged from two epoches;
$^{11)}$ -- erros assumed are the same as in Asplund et al.\,(\cite{AGL00}).
\item[~~]
\end{list}
\end{table*}
\begin{table*}[t]{}
	\centerline{Table\,\ref{table1}. (continued)}
\begin{list}{}{}
\item[The abbreviations used in column (10):]
\end{list}
\centering
\begin{tabular}{l@{ }l@{  }l@{  }}
AFGM01: Arellano Ferro et al.\,(\cite{AFGM01})&   KDK97:  Kendall et al.\,(\cite{KDK97})       & RGA90:   Rao et al.\,(\cite{RGA90})              \\        
AGL00:  Asplund et al.\,(\cite{AGL00})        &   KL86:   Kilkenny \& Lydon\,(\cite{KL86})     & RGL02:   Rao et al.\,(\cite{RGL02})              \\
CDK94:  Conlon et al.\,(\cite{CDK94})         &   K95:    Klochkova\,(\cite{K95})              & RL03:    Rao \& Lambert\,(\cite{RL03})           \\
CDK91:  Conlon et al.\,(\cite{CDK91})         &   KP96:   Klochkova \& Panchuk\,(\cite{KP96})  & RBH99:   Reddy et al.\,(\cite{RBH99})            \\
CTM93:  Conlon et al.\,(\cite{CTM93})         &   KPT02:  Klochkova et al.\,(\cite{KPT02})     & RH99:    Reddy \& Hrivnak\,(\cite{RH99})         \\
DJH98:  Drilling et al.\,(\cite{DJH98})       &   KSP00a: Klochkova et al.\,(\cite{KSP00a})    & RLG02:   Reddy et al.\,(\cite{RLG02})            \\
DvWW98: Decin et al.\,(\cite{DvWW98})         &   KSP00b: Klochkova et al.\,(\cite{KSP00b})    & RPS96:   Reddy et al.\,(\cite{RPS96})            \\
GP03:   Gauba \& Parthasarathy\,(\cite{GP03}) &   KSP99:  Klochkova et al.\,(\cite{KSP99})     & RDvW04:  Reyniers et al.\,(\cite{RDvW04})        \\
GAFP97: Giridhar et al.\,(\cite{GAFP97})      &   KYM02:  Klochkova et al.\,(\cite{KYM02})     & RvW01:   Reyniers \& Van Winckel\,(\cite{RvW01}) \\
GLG98:  Giridhar et al.\,(\cite{GLG98})       &   K73:    Kodaira\,(\cite{K73})                & RvWG04:  Reyniers et al.\,(\cite{RvWG04})        \\
GLG00:  Giridhar et al.\,(\cite{GLG00})       &   LB84:   Luck \& Bond\,(\cite{LB84})          & RDM03:   Ryans et al.\,(\cite{RDM03})            \\
GRL94:  Giridhar et al.\,(\cite{GRL94})       &   LBL90:  Luck et al.\,(\cite{LBL90})          & SPG99:   Sivarani et al.\,(\cite{SPG99})         \\
GLG97a: Gonzalez et al.\,(\cite{GLG97a})      &   LLB83:  Luck et al.\,(\cite{LLB83})          & TPJ00:   Th{\'e}venin et al.\,(\cite{TPJ00})    \\
GLG97b: Gonzalez et al.\,(\cite{GLG97b})      &   MvWW02: Maas et al.\,(\cite{MvWW02})         & vW95:    Van Winckel\,(\cite{vW95})              \\
GW92:   Gonzalez \& Wallerstein\,(\cite{GW92})&   MvWLE05: Maas et al.\,(\cite{MvWLE05})       & vW97:    Van Winckel\,(\cite{vW97})              \\
HDK96:  Hambly et al.\,(\cite{HDK96})         &   MCD92:  McCausland et al.\,(\cite{MCD92})    & vWOT96:  van Winckel et al.\,(\cite{vWOT96})     \\
HJ97:   Harrison \& Jeffery\,(\cite{HJ97})    &   MH98:   Moehler \& Heber\,(\cite{MH98})      & vWR00:   Van Winckel \& Reyniers\,(\cite{vWR00}) \\
H83:    Heber\,(\cite{H83})                   &   MRK02:  Mooney et al.\,(\cite{MRK02})        & vWWW96:  Van Winckel et al.\,(\cite{vWWW96})     \\
HR03:   Hrivnak \& Reddy\,(\cite{HR03})       &   MDS04:  Munn et al.\,(\cite{MDS04})          & VSL98:   Veen et al.\,(\cite{VSL98})             \\
J93:    Jeffery\,(\cite{J93})                 &   NHK94:  Napiwotzki et al.\,(\cite{NHK94})    & WvWB91:  Waelkens et al.\,(\cite{WvWB91})        \\
JH92:   Jeffery \& Heber\,(\cite{JH92})       &   PGS00:  Parthasarathy et al.\,(\cite{PGS00}) & WvWTW92: Waelkens et al.\,(\cite{WvWTW92})       \\
JH93:   Jeffery \& Heber\,(\cite{JH93})       &   PLM04:  Pereira et al.\,(\cite{PLM04})       & WS81:    Walker \& Sch{\"o}nberner\,(\cite{WS81})\\
JHHJ98: Jeffery et al.\,(\cite{JHHJ98})       &   PRL01:  Pandey et al.\,(\cite{PRL01})        & ZKP95:   Za\'cs et al.\,(\cite{ZKP95})           \\  
JHH99:  Jeffery et al.\,(\cite{JHH99})        &   PLRJ04: Pandey et al.\,(\cite{PLRJ04})       & ZKP96:   Za\'cs et al.\,(\cite{ZKP96})           \\
\end{tabular}
\end{table*}

%********************************* table 2 *************************
\begin{table*}[t]{}
\caption[]{Chemical composition of our  of post-AGB sample.}
\label{table2}
\centering
\centering
\begin{tabular}{l@{ }l@{  }r@{  }r@{  }r@{  }r@{  }r@{  }r@{  }}
\hline
\noalign{\smallskip}
 No & name                & ~$\epsilon$(C)~ & ~$\epsilon$(N)~ & ~$\epsilon$(O)~ & ~$\epsilon$(S)~  & ~$\epsilon$(Fe)~ &  ~$\epsilon$(Zn)~  \\  
(1) &(2)                  &(3)            &(4)            &(5)            &(6)             &(7)             & (8)              \\
\hline
\noalign{\smallskip}
  1 &  006.72 $-$10.37    & 7.93          & 7.34          & 8.29          & 7.17           & 6.89           &  4.37            \\
    &                     & 8.01          & 7.61          & 8.47          & 7.06           & 6.94           &  4.17            \\
  2 &  007.96 $+$26.71    & 8.35          & 8.55          & 8.85          & 7.30           & 7.10           &  4.76            \\
  3 &  013.23 $+$12.17    & 8.58          & 7.65          & 8.41          & 7.22           & 7.12           &  4.52            \\
    &                     & 8.30          &               & 8.33          & 7.29           & 6.92           &                  \\
  4 &  016.45 $-$50.43    &$<$6.80        & 7.00          & 8.00          &                &                &                  \\
  5 &  023.98 $-$21.04    & 8.98          & 8.37          & 8.96          & 6.86           & 6.91           &                  \\
    &                     & 9.08          & 8.32          & 9.13          & 7.36           & 7.17           &                  \\
  6 &  029.18 $-$21.26    & 7.50          & 7.65          & 8.51          & 7.40           &                &                  \\
    &                     & 6.70          & 7.80          & 8.80          & 6.60           & 6.70           &                  \\
  7 &  030.60 $-$21.53    & 7.03          & 7.11          & 7.51          & 6.32           & 6.38           &  3.80            \\
  8 &  031.33 $-$43.48    & 6.40          & 7.50          & 8.20          & 6.70           & 6.20           &                  \\
    &                     &               &               &               &                &                &                  \\
  9 &  033.16 $-$48.12    &$<$6.10        & 6.80          & 7.90          &$<$6.50         &                &                  \\
 10 &  035.62 $-$04.96    & 8.74          & 9.10          & 9.37          & 7.38           & 7.35           &                  \\
    &                     & 9.17          &               &               &                & 7.37           &                  \\
    &                     &$<$7.52        &               &               &                &$<$7.00         & $<$4.35          \\
 11 &  040.51 $-$10.09    & 7.74          &               &               &                & 6.40           &  4.07            \\
 12 &  043.06 $+$32.36    & 6.80          &$<$6.70        &               &                &                &                  \\
    &                     & 7.01          & 7.02          & 7.90          &                &                &                  \\
 13 &  043.23 $-$57.13    & 9.02          & 8.69          & 8.76          & 7.57           &  6.78          &                  \\
 14 &  050.67 $+$19.79    & 7.19          & 7.43          & 8.47          &                &                &                  \\
    &                     & 7.12          & 7.47          & 8.36          &                &                &    .             \\
    &                     & 6.92          & 7.22          & 8.23          & 6.33           & 6.90           &                  \\
    &                     & 7.04          & 7.52          & 8.35          & 7.25           &                &                  \\
 15 &  051.43 $+$23.19    & 8.30          & 8.29          & 8.66          & 7.01           & 7.08           &  4.27            \\
 16 &  052.73 $+$50.79    & 7.40          & 7.30          & 8.10          &                & 5.50           &                  \\
 17 &  053.84 $+$20.18    & 8.27          & 7.66          & 8.74          & 6.96           & 6.72           &  4.60            \\
 18 &  066.18 $+$18.58    & 7.25          &               & 9.20          & 6.88           &  6.89          &                  \\
 19 &  067.16 $+$02.73    & 8.46          & 9.02          & 9.29          & 7.37           &  7.27          &  4.41            \\
    &                     & 8.22          &               & 8.93          & 7.50           &  7.27          &                  \\
 20 &  077.13 $+$30.87    & 8.45          & 8.92          & 9.01          & 7.68           &  7.17          &  4.89            \\
    &                     & 8.52          & 8.39          & 9.15          & 7.46           &  7.26          &  4.63            \\
 21 &  080.17 $-$06.50    & 8.69          & 9.38          & 8.68          & 7.09           &  6.92          &  3.91            \\
 22 &  096.75 $-$11.56    & 8.58          & 7.84          & 8.50          & 6.95           &  7.20          &                  \\
    &                     & 8.63          & 7.88          & 8.50          & 6.85           &  7.07          &                  \\
 23 &  098.41 $-$16.73    & 8.20          & 8.30          & 8.80          & 7.30           &  4.55          &                  \\
 24 &  103.35 $-$02.52    & 8.69          & 7.68          & 8.48          & 6.89           &  6.69          &  4.16            \\
    &                     & 7.37          &               & 8.77          &                &  7.02          &  6.30            \\
 25 &  113.86 $+$00.59    & 8.70          & 7.68          & 8.24          & 6.98           &  6.72          &                  \\
    &                     & 8.89          & 8.69          & 9.03          & 7.05           &  6.86          &  4.68            \\
 26 &  123.57 $+$16.59    & 8.32          & 7.70          & 8.24          &                &  7.19          &                  \\
 27 &  133.73 $+$01.50    & 8.84          & 8.67          &               & 7.07           &  7.03          &                  \\
 28 &  161.98 $+$19.59    & 8.19          &               & 8.47          &                &  5.95          &  3.24            \\
 29 &  166.24 $-$09.05    & 8.71          & 7.76          &               & 7.02           &  6.89          &                  \\
    &                     & 8.55          & 7.96          & 8.22          & 6.80           &  6.66          &  3.84            \\
    &                     & 8.81          & 7.84          &               & 7.13           &  6.82          &                  \\
 30 &  172.95 $-$05.50    &$<$6.70        & 7.30          & 7.60          &$<$6.29         & $<$6.70        &                  \\
 31 &  173.86 $-$82.41    &$<$6.90        & 7.40          & 8.20          &$<$6.50         & $<$6.20        &                  \\
 32 &  188.86 $-$14.29    & 8.81          & 8.24          & 8.43          & 6.87           &  6.75          &  3.74            \\
 33 &  196.19 $-$12.14    & 8.73          & 7.83          & 8.57          & 6.64           &  6.66          &                  \\
 34 &  204.67 $+$07.57    & 8.14          & 7.55          & 8.47          & 6.19           &  3.05          &                  \\
 35 &  206.75 $+$09.99    & 8.09          & 7.87          & 8.40          & 6.49           &  6.50          &                  \\
    &                     & 8.65          & 7.84          & 8.67          & 6.61           &  6.51          &  3.97            \\
    &                     & 8.63          & 8.00          & 8.49          & 6.84           &  6.50          &                  \\
 36 &  208.93 $+$17.07    & 8.76          & 7.97          &               & 6.98           &  7.06          &                  \\
 37 &  215.44 $-$00.13    & 8.72          & 7.97          & 8.30          &                &  6.59          &                  \\
    &                     & 9.09          &               & 8.64          & 6.97           &  7.03          &                  \\
 38 &  218.97 $-$11.76    & 8.62          & 7.82          & 8.72          & 7.04           &  4.00          &                  \\
  \noalign{\smallskip}
  \hline
\end{tabular}
\end{table*}
\begin{table*}[t]{}
         	\centerline{Table\,\ref{table2}. (continued)}
\centering
\begin{tabular}{l@{ }l@{  }r@{  }r@{  }r@{  }r@{  }r@{  }r@{  }}
\hline
\noalign{\smallskip}
 No & name                & ~$\epsilon$(C)~ & ~$\epsilon$(N)~ & ~$\epsilon$(O)~ & ~$\epsilon$(S)~  & ~$\epsilon$(Fe)~ &  ~$\epsilon$(Zn)~  \\  
(1) &(2)                  &(3)            &(4)            &(5)            &(6)             &(7)             & (8)              \\
\hline
\noalign{\smallskip}
 39 &  236.57 $-$05.37    & 7.94          & 7.56          &               & 6.39           &  6.67          &  3.60            \\
 40 &  260.83 $-$05.07    & 8.64          & 7.76          & 8.52          & 7.07           &  7.12          &  4.54            \\
 41 &  264.55 $+$72.47    & 8.46          & 7.79          & 8.37          &                &  6.30          &                  \\
 42 &  266.85 $+$22.93    & 8.41          & 8.07          & 8.64          & 6.82           &  2.86          &                  \\
 43 &  269.97 $-$34.08    & 8.25          & 8.24          & 8.82          & 6.90           &  5.82          &                  \\
 44 &  290.54 $-$01.95    & 8.51          & 7.98          & 8.52          & 7.09           &  7.61          &  4.29            \\
 45 &  293.03 $+$05.94    & 9.55          & 8.50          & 8.97          &                &  7.50          &                  \\
 46 &  295.48 $+$29.87    & 7.11          & 7.31          & 7.81          & 6.14           &  6.30          &                  \\
 47 &  298.25 $+$15.48    & 8.54          & 8.16          & 8.80          & 6.97           &  7.43          &  4.31            \\
 48 &  298.30 $+$08.67    & 8.11          & 7.76          & 8.70          & 7.15           &  6.77          &  3.78            \\
 49 &  299.02 $-$43.77    & 6.70          & 7.60          & 7.60          & 5.90           & $<$6.70        &                  \\
 50 &  304.34 $+$36.40    & 7.28          &               & 8.59          & 6.53           &  6.42          &  3.76            \\
    &                     & 7.52          & 7.34          & 8.57          & 6.15           &  6.31          &                  \\
 51 &  309.07 $+$15.18    & 7.46          & 7.30          & 8.17          & 6.53           &  5.70          &  3.44            \\
 52 &  317.11 $+$53.11    &$<$6.80        & 6.80          &$<$7.90        &                &                &                  \\
    &                     &$<$6.16        & 6.93          & 7.98          &                &                &                  \\
 53 &  325.04 $+$08.65    & 7.94          & 7.78          & 8.52          & 6.83           &  6.81          &                  \\
 54 &  326.77 $-$07.49    &               &               &               &                &                &                  \\
    &                     &               & 8.00          & 9.30          &                &  6.70          &                  \\
 55 &  330.64 $-$03.67    &               & 7.20          & 8.30          &                &  5.50          &                  \\
 56 &  345.58 $-$07.30    & 7.83          & 7.57          & 8.60          &                &  7.56          &                  \\
\noalign{\smallskip}
\multicolumn{7}{l}{\textbf{RV Tau stars}}
\\
\noalign{\smallskip}
 57 &  020.72 $+$17.57    & 6.61          &               & 6.93          &                &  5.34          &                  \\
    &                     & 6.08          &$<$6.50        & 7.62          &                &  5.29          &  2.50            \\
 58 &  021.48 $+$28.63    & 8.07          &               & 8.33          & 7.25           &  6.65          &  3.91            \\
 59 &  026.52 $-$05.42    & 8.24          &               & 8.63          & 7.21           &  5.38          &  4.50            \\
 60 &  044.10 $-$61.56    & 7.26          &               & 8.53          & 6.42           &  6.36          &  3.55            \\
    &                     &               &               & 8.54          & 6.16           &  6.50          &                  \\
    &                     & 6.80          &$<$6.60        & 7.88          &                &  5.90          &  3.11            \\
 61 &  050.49 $+$14.24    & 8.50          &               & 8.65          & 6.90           &  6.10          &  3.69            \\
 62 &  057.53 $-$09.75    & 8.15          &               & 8.29          & 7.58           &  7.01          &  4.41            \\
 63 &  058.44 $-$07.46    & 8.85          & 8.83          & 9.15          & 7.47           &  7.24          &  4.46            \\
 64 &  060.73 $+$06.94    & 8.19          & 7.82          & 8.83          & 6.60           &  5.71          &  3.90            \\
 65 &  076.75 $-$11.78    &$<$6.15        &               & 8.34          & 6.36           &  6.10          &  3.26            \\
 66 &  148.26 $+$05.26    & 8.79          &               & 8.44          & 7.19           &  7.00          &  4.28            \\
 67 &  174.77 $-$12.19    & 9.23          &               & 8.69          &                &  7.07          &  4.64            \\
 68 &  188.06 $+$01.30    & 8.15          & 7.39          & 8.38          & 6.96           &  6.63          &  4.59            \\
 69 &  195.41 $-$03.42    & 8.38          &               & 8.94          & 7.37           &  5.20          &  4.81            \\
 70 &  199.39 $-$04.56    & 8.02          &               & 8.29          & 6.85           &  5.63          &  3.99            \\
 71 &  217.80 $+$09.95    &               &               &               &                &  6.32          &  4.17            \\
 72 &  226.14 $+$04.15    & 8.37          &               & 8.46          & 7.09           &  6.71          &  3.91            \\
 73 &  253.02 $-$03.00    & 8.60          &               & 9.01          & 7.65           &  6.64          &                  \\
 74 &  282.42 $-$09.24    & 8.87          &               & 8.54          & 7.59           &  6.45          &  4.60            \\
 75 &  293.17 $-$17.24    & 8.32          &               & 9.13          & 7.03           &  6.90          &  4.10            \\
 76 &  295.25 $+$16.82    & 8.14          & 7.72          & 8.57          & 6.53           &  5.63          &  3.60            \\
 77 &  297.87 $+$13.36    & 8.50          & 8.42          & 9.00          & 7.13           &  6.37          &  4.06            \\
 78 &  313.90 $-$08.68    & 8.21          & 7.83          & 8.29          & 6.65           &  7.00          &  4.14            \\
 79 &  330.84 $+$57.77    &               &               &               &                &  6.31          &  3.87            \\
 80 &  331.86 $-$13.78    & 8.39          &               &               & 7.25           &  6.48          &  4.33            \\
 81 &  339.79 $-$04.68    & 8.82          &               &               &                &  6.00          &  3.40            \\
 82 &  344.12 $+$26.45    & 7.30          &               & 8.53          & 6.48           &  6.32          &  3.56            \\
 83 &  345.49 $-$04.99    & 8.32          & 7.72          & 8.72          & 7.33           &  7.50          &  4.80            \\
 84 &  345.54 $+$10.26    & 8.02          & 7.42          & 8.43          & 7.03           &  6.80          &  4.20            \\
 85 &  345.65 $-$04.69    & 8.32          & 8.12          & 8.53          & 7.43           &  6.50          &  4.30            \\
  \noalign{\smallskip}
  \hline
\end{tabular}
\end{table*}
\begin{table*}[t]{}
         	\centerline{Table\,\ref{table2}. (continued)}
\centering
\begin{tabular}{l@{ }l@{  }r@{  }r@{  }r@{  }r@{  }r@{  }r@{  }}
\hline
\noalign{\smallskip}
 No & name                & ~$\epsilon$(C)~ & ~$\epsilon$(N)~ & ~$\epsilon$(O)~ & ~$\epsilon$(S)~  & ~$\epsilon$(Fe)~ &  ~$\epsilon$(Zn)~  \\   
(1) &(2)                  &(3)            &(4)            &(5)            &(6)             &(7)             & (8)              \\
\hline
\noalign{\smallskip}
\multicolumn{7}{l}{\textbf{suspected RV Tau stars from Maas et al.\,\cite{MvWLE05}}}
\\
\noalign{\smallskip}
 86 &   33.59 $-$07.22    & 8.42          &               & 9.02          &                &  7.60          &                  \\
 87 &   39.02 $-$03.49    & 8.72          &               & 9.03          & 7.53           &  7.20          &  4.70            \\
 88 &  254.58 $+$12.94    & 7.62          &               &               & 6.53           &  6.80          &  4.00            \\
 89 &  265.50 $+$00.39    & 8.22          & 7.82          & 8.33          & 7.53           &  7.20          &  4.70            \\
 90 &  271.51 $-$00.50    & 8.22          & 8.12          & 8.53          &                &  7.20          &                  \\
 91 &  327.82 $+$00.63    & 8.82          & 8.52          & 8.83          & 7.93           &  7.50          &  4.90            \\
\noalign{\smallskip}
\multicolumn{7}{l}{\textbf{R CrB stars}}
\\
\noalign{\smallskip}
 92 &  002.41 $+$07.51    & 9.11          & 8.42          & 8.60          & 6.70           &  6.40          &  4.16            \\
 93 &  002.52 $-$05.97    & 8.80          & 7.60          & 8.70          & 6.70           &  5.80          &  3.90            \\
 94 &  004.43 $-$19.45    & 8.90          & 8.50          & 7.90          & 7.30           &  6.70          &  4.50            \\
 95 &  005.81 $-$03.78    & 8.80          & 8.00          & 7.50          & 7.40           &  5.60          &  4.10            \\
 96 &  008.31 $-$05.24    & 8.80          & 8.70          & 8.20          & 7.00           &  6.30          &  4.40            \\
 97 &  023.83 $-$02.92    & 8.80          & 8.79          & 7.70          & 7.00           &  6.30          &  4.10            \\
 98 &  045.05 $+$50.98    & 9.20          & 8.40          & 9.00          & 6.80           &  6.50          &                  \\
    &                     &               &               &               &                &                &                  \\
 99 &  070.45 $+$02.20    & 8.90          & 8.80          & 8.10          & 6.90           &  6.70          &  4.40            \\
100 &  109.52 $-$00.39    & 9.20          & 8.50          & 7.50          & 7.00           &  6.90          &  4.80            \\
101 &  149.84 $+$01.12    & 9.00          & 8.90          & 8.40          & 6.80           &  6.80          &                  \\
102 &  188.86 $-$04.42    & 8.80          & 8.50          & 8.40          & 6.50           &  6.10          &  3.60            \\
103 &  279.06 $+$20.12    & 8.90          & 8.30          & 8.80          & 6.20           &  6.20          &                  \\
104 &  301.74 $+$08.32    & 8.60          & 8.30          & 7.70          & 6.70           &  6.30          &  4.30            \\
105 &  304.42 $-$02.68    & 8.90          & 8.80          & 7.70          & 6.90           &  6.50          &  4.40            \\
106 &  307.96 $+$08.29    & 9.51          & 8.01          & 8.85          & 7.11           &  5.04          &                  \\
107 &  327.22 $-$06.92    & 8.90          & 9.10          & 8.40          & 7.70           &  6.80          &  4.70            \\
108 &  332.44 $-$03.57    & 8.90          & 8.70          & 8.90          & 6.80           &  6.60          &  4.40            \\
109 &  347.53 $-$14.14    & 8.90          & 8.80          & 8.30          & 6.80           &  6.40          &  4.30            \\
110 &  357.66 $-$15.65    & 8.60          & 8.60          & 8.70          & 7.50           &  5.50          &  2.90            \\
\noalign{\smallskip}
\multicolumn{7}{l}{\textbf{extreme helium stars}}
\\
\noalign{\smallskip}
111 &  006.01 $+$26.02    & 9.17          & 7.97          & 7.90          & 7.83           &  6.57          &                  \\
112 &  007.82 $+$05.07    & 9.38          & 8.16          & 9.39          & 7.12           &  6.84          &                  \\
113 &  020.91 $-$08.31    & 9.45          & 8.60          & 8.50          & 7.65           &  6.90          &                  \\
    &                     & 8.90          & 8.60          &               & 7.10           &  6.30          &                  \\
114 &  024.41 $+$12.50    & 9.30          & 8.25          & 8.85          & 6.70           &  6.30          &                  \\
115 &  026.55 $+$10.09    & 9.00          & 8.50          & 8.40          & 6.90           &  6.70          &                  \\
116 &  031.33 $+$33.28    &               & 8.93          & 7.54          & 7.44           &  7.40          &                  \\
117 &  049.87 $-$25.21    & 9.00          & 7.15          & 8.90          & 6.00           &  5.40          &                  \\
    &                     & 8.60          & 7.20          & 8.40          & 5.80           &  5.50          &  3.60            \\
118 &  068.90 $+$04.76    & 9.65          & 8.60          & 9.60          & 7.20           &  6.90          &  4.60            \\
119 &  222.95 $-$04.18    & 9.13          & 7.61          & 8.59          & 6.92           &  6.89          &                  \\
120 &  235.21 $+$54.44    & 9.54          & 8.11          & 8.10          & 7.12           &  6.49          &                  \\
121 &  309.95 $-$04.25    & 9.02          & 8.26          & 8.05          & 6.67           &  6.52          &                  \\
122 &  317.65 $+$14.18    & 9.40          &               &               &                &  7.10          &  4.20            \\
123 &  338.13 $-$18.71    & 9.50          & 8.90          & 8.40          & 7.00           &  7.50          &                  \\
124 &  344.19 $-$08.84    & 9.54          & 8.33          & 9.06          & 7.13           &  6.78          &                  \\
125 &  348.17 $+$17.78    & 6.80          & 6.46          & 6.95          &                &  6.38          &                  \\
	\noalign{\smallskip}
	\hline
\end{tabular}

\end{table*}

\begin{figure*}[htbp]
\resizebox{\hsize}{!}{\includegraphics{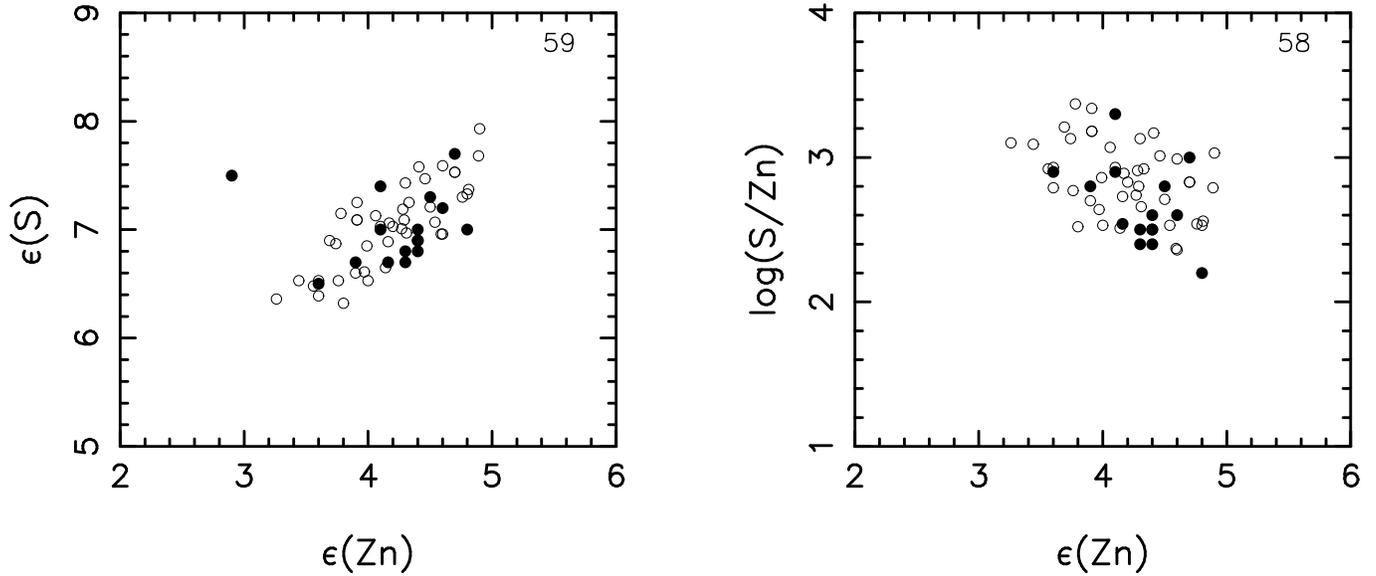}}
       \caption{{\bf a}-(left) The $\epsilon$(S) vs. $\epsilon$(Zn) relation in our post-AGB sample. {\bf b}-(right) The S/Zn vs. $\epsilon$(Zn) relation. Objects from 
the R CrB class and extreme helium stars are represented by black 
circles. The number of objects in our sample with available data is indicated 
in the top right corner of each panel.}
       \label{fig1}
\end{figure*}

Table 2 is ordered in the same way as Table 1.  Its  columns contain: (1) the object number;  (2) the object coordinates {\it l}, {\it b};
(3) -- (8) the abundances of C, N, O, S, Fe, and Zn,  expressed as $\epsilon$(X)$\,=\,$12 + log (X/H)\, where X/H is the abundance of element X 
in number relative to H.  In the case of R\,CrB and extreme helium stars, the abundances 
are listed as $\epsilon$(X)\,=\,12.15 + log(X/$N$), where X/$N$ is the ratio of 
number density of element X to the total number density of nucleons $N$. Note 
that the second definition is more general than the earlier one and both are 
consistent if the abundance of helium amounts to $\epsilon$(He)\,=\,11, a condition which is fulfilled with good accuracy in the remaining stars (see Asplund et al.\,\cite{AGL00} and  Pandey\,et al.\,\cite{PRL01}).
The sources for the abundances are the same as given 
in col. (8) of Table 1. Overall, the 
typical uncertainty in elemental abundances is about 0.2-0.3 dex. However, 
the uncertainty in abundance ratios of heavy elements is smaller, since many 
sources of errors affect the derived abundances in a similar way. One exception is the case of oxygen, if its abundance has been determined from the O~{\sc i}  7771-5 triplet. It is well known that this triplet gives enhanced abundances, if non-LTE effects are not taken into account, and that the O~{\sc i} 7771-5 vs [O~{\sc i} ] discrepancy is higher for low metallicity (Takeda\,\cite{Takeda_2003}). In most objects listed in Table 2, it appears that the use of this triplet was avoided.

\subsection{Notes on individual objects and discussion of abundance 
uncertainties}
\label{notes}

For several sources we found more than one reference with stellar 
parameters and chemical composition determined. Below we present arguments for 
the preferred source of information. The object number corresponds to the 
number given in column (1) of Tables\,1 and 2. 

\noindent {\bf Object 1: IRAS\,18384$-$2800.} The atmospheric parameters and chemical 
abundances have been analyzed recently by AFGM01 and RvW01. The 
same atmospheric parameters were derived in both analyses. The spectra obtained by RvW01 have 
apparently much higher S/N-ratio than those of AFGM01, and the chemical analysis of RvW01 
is based on a higher number of lines, so the results of RvW01 were chosen for 
further analysis. 

\noindent {\bf Object 3: IRAS\,17279$-$1119.} This star has been analyzed by AFGM01 and VW97. Both 
analyses gave similar value for $T_{\rm eff}$ but rather different values for the 
surface gravity. Since the determination of AFGM01 was based on a single 
spectrum, while the vW97 determination was based on several spectra at different 
photometric phases, it is likely that  more consistent atmospheric parameters 
and chemical composition are obtained in the AFGM01 paper.

\noindent {\bf Object 5: IRAS\,19500$-$1709.}  Both analyses by vWR00 and vWWW96 give 
similar atmospheric parameters. The vWR00 paper is based on a higher S/N and a 
broader spectral coverage and was chosen for the subsequent analysis.

\noindent {\bf Object 6: IRAS\,19590$-$1249.} We adopted 
the RDM03 values. The analysis of RDM03 is based on higher quality material and 
is based on fully blanketed non-LTE atmospheric models that
should guarantee more accurate values for $T_{\rm eff}$ and log\,$g$.

\noindent {\bf Object 8: PHL\,1580.} KL86 do not give abundances, so we had to 
rely on those given by CDK91.

\noindent {\bf Object 10: IRAS\,19114$+$0002.} The analysis of RH99 is based on higher 
quality material than in ZKP96. As a consequence the microturbulence 
derived by RH99 (5.25 km s$^{-1}$) is much lower than the supersonic value 
found by ZKP99 (8 km s$^{-1}$). Since TPJ00 give a really unusual result we 
selected the data from RH99.

\noindent {\bf Object 12: PG\,1704$+$222.} The analysis of MH98 is based on better 
quality spectra than the preliminary results of CTM93, and their results were 
adopted.

\noindent {\bf Object 14: IRAS\,18062$+$2410.} There are emission lines visible in the 
spectrum. The four papers containing a determination of atmospheric parameters 
are not independent. The paper by AFGM01 was suggested by the work of PGS00. It is 
not quite clear how the atmospheric parameters were determined. The analysis of 
RDM03 is based on non-LTE, fully blanketed atmospheric models and the same 
spectroscopic material as used in the LTE analysis by MRK02. Hence preference 
is given to work by RDM03. Their values consistently point to a higher mass of 
the star. 

\noindent {\bf Object 19: IRAS\,19475$+$3119.} Since the paper of KPT02 is based on spectra 
with relatively low resolution (15\,000), the analysis by AFGM01 is probably 
more reliable.

\noindent {\bf Object 20: IRAS\,17436$+$5003.} Both analysis by KPT02 and LBL90 result in a 
massive post-AGB star. KPT02 spectra have lower resolution than those of LBL90, 
so we adopted the data from the latter.

\noindent {\bf Object 22: IRAS\,22223$+$4327.} Both papers by vWR00 and DvWW98 are based on 
the same spectroscopical material and the same methods of analysis. We adopted 
the data from the more recent paper vWR00.

\noindent {\bf Object 24: IRAS\,22272$+$5435.} Both papers by ZKP95 and RLG02 give consistent 
atmospheric parameters but differ in derived abundances, particularly of Fe. 
Other determinations are based on relatively few lines (C, N) or on just one line 
(Zn, O) and may be in error. The paper by RLG02 is based on higher quality 
material so we used their results.

\noindent {\bf Object 25: IRAS\,23304$+$6147} The vWR00 analysis is based on much 
higher resolution spectra (60\,000) compared to that of KSP00 (15\,000), 
so we used the results from vWR00.

\noindent {\bf Object 29: IRAS\,04296$+$3429.} The works by vWR00 and DvWW98 are 
based on the same spectroscopical material and the same methods of analysis. 
Data from vWR00 were adopted. KSP99 spectra have lower resolution.

\noindent {\bf Object 35: IRAS\,07134$+$1005.} All analyses of HD\,56126 give similar values 
of atmospheric parameters. The papers by HR03 and vWR00 are based on a higher 
quality spectrum than the paper by K95, yet there is a difference in the 
absolute determination of the carbon abundance. The remaining abundances are 
similar. Data from vWR00 were adopted.

\noindent {\bf Object 37: IRAS\,06530$-$0213.} The analysis of RvWG04 is based on 
higher quality spectroscopic material than  HR03, so their results were 
used.

\noindent {\bf Object 50: IRAS\,12538$-$2611.} The analysis of GAFP97 is based on higher 
quality spectra and more lines are used for the determination of the chemical 
abundances, so  their values were preferred.

\noindent {\bf Object 52: BPS\,CS\,22877$-$0023.} Both analyses lead to uncertain 
determinations of atmospheric parameters, but MH98 used a larger number of 
methods, so we adopted their values. 

\noindent {\bf Object 54: LS 3591.} We adopted the VSL98 results since 
only their work gives chemical abundances.

\noindent {\bf Object 57: V453\,Oph and object 60: HD\,216457.} The abundance analysis of 
RDvW04 is based on higher quality spectra and a wider spectral
range than that of GLG00 and GLG98, so we adopted the results from RDvW04.

\noindent {\bf Object 98: IRAS\,15465$+$2818.} The paper by AGL00 is the only one that gives 
chemical abudances.

\noindent {\bf Object 113: V4732\,Sgr and object 117: FQ Aqr.} The paper of PRL01 gives a 
consistent set of abundances for the sample of extreme helium stars
analyzed here, so their results are preferred.

\section{Empirical diagrams}

\subsection{Choice of a metallicity indicator}

Since it is expected that yields are strongly dependent on ``metallicity'' or, 
better said, on the initial chemical composition of the star, we first have 
to choose a reasonable metallicity indicator. As stressed, e.g., by	
Mathis \& Lamers\,(\cite{Mathis_and_Lamers_1992}) or Lambert\,(\cite{Lambert_2004}),
the usual metallicity indicator in stellar atmospheres, Fe, cannot be used for post-AGB stars because of 
possible strong depletion in dust grains  in a former stage and subsequent 
ejection of the grains (dust-gas separation). 
Oxygen,  the most abundant `'metal'' and thus the best theoretical  metallicity indicator,  is possibly affected by nucleosynthesis on the 
AGB (ON cycle, hot bottom burning). From the list of elements for which we 
compiled the abundances, good metallicity indicators would be S and Zn.  The first one is an $\alpha$-element, like O, and its abundance in the Galaxy is proportional to that of O. For the second one, the nucleosynthesis  mechanism  is unknown a priori, but Zn appears to roughly follow Fe (Mishenina et al. 2002). Figure 
1a shows $\epsilon$(S) vs. $\epsilon$(Zn) in our objects. The correlation is quite good (there is one outlier:  V CrA, which also appears to be as an extreme object in many of the abundance-ratio diagrams of Asplund et al. 2000). The dispersion gives an 
idea of a realistic average abundance uncertainty: about 0.3 dex. Note that the data 
set spans a metallicity range of 2 dex.  Figure 1b shows that S/Zn has  a tendency to increase  as $\epsilon$(Zn) decreases (on average by 0.5\,dex per dex). This is the well-known  $\alpha$-enhancement observed in Population II stars (Norris et al. 2001). The effect of  $\alpha$-enhancement on stellar parameters and stellar evolution is complex (Kim et al. 2002) and has not yet been investigated in AGB stars. We chose S as our principal 
metallicity indicator, since there are more stars with determinations of S 
abundances (113) than of Zn abundances (76). We keep Zn as a secondary indicator that might test the effect of $\alpha$-enhanced mixtures in the stars. We note, however, that the Zn abundance measurements often rely only on one  line, which makes the evaluation of statistical errors difficult. 

Figure 2 shows $\epsilon$(O) vs. $\epsilon$(S). The dispersion is such that obviously oxygen cannot 
be chosen as a metallicity indicator in our sample. The reason for this 
dispersion is not clear a priori. It can be due to nucleosynthesis and mixing 
affecting the oxygen abundance. But it could also be due to larger uncertainties in the oxygen 
abundance than was thought. Note that the R Cr B stars and extreme helium stars show the largest dispersion   and the larger proportion of objects with O/S smaller than solar.

\subsection{Nitrogen enhancement}

\begin{figure}[]
\resizebox{\hsize}{!}{\includegraphics{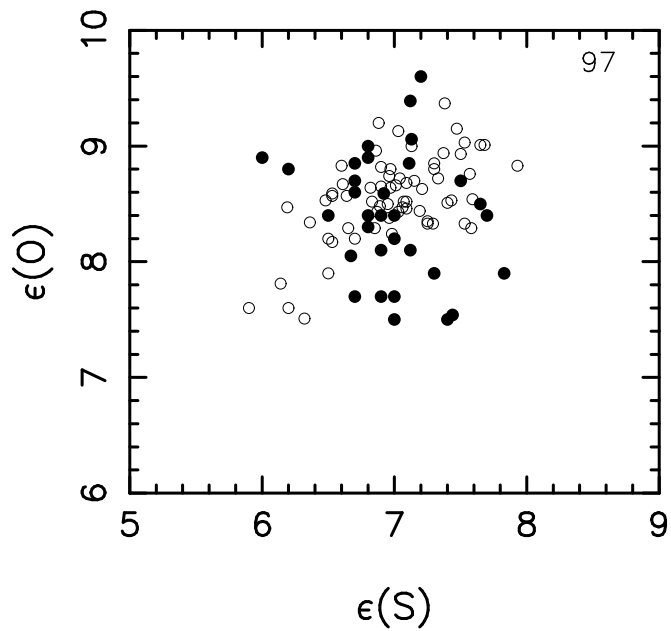}}
\caption{ The $\epsilon$(O) vs. $\epsilon$(S) relation.}
       \label{fig2}
\end{figure}

Figure 3 shows log N/O vs. $\epsilon$(O).  The objects from the R CrB class and the extreme 
helium stars (which are represented by black circles) show 
different behaviour from the rest: they draw a clear anticorrelation between 
N/O and $\epsilon$(O). This diagram has also been constructed for planetary nebulae (e.g. Henry 
et al.\,\cite{Henry_etal_1989}, Kingsburgh \& Barlow\,\cite{Kingsburgh_and_Barlow_1994}, 
Leisy \& Dennefeld\,\cite{Leisy_and_Dennefeld_1996}) with different 
results depending on the authors and on the samples.  Some claim not to see 
any anticorrelation. To our knowledge, never has the anticorrelation been seen 
so prominently for a class of PNe as for our R CrB and extreme helium stars
subsample of post-AGB stars. One interpretation of such an anticorrelation is 
the production of N at the expense of O (ON cycle) brought to the stellar 
surface by the 2nd dredge-up.  However, the N/O values reached by R CrB stars are much higher than for the remaining post-AGB stars and for planetary nebulae.  Asplund et al. (2000) argue that CNO cycling on He-burning products
has to be invoked to reach this high N enhancement.

\begin{figure}[]
\resizebox{\hsize}{!}{\includegraphics{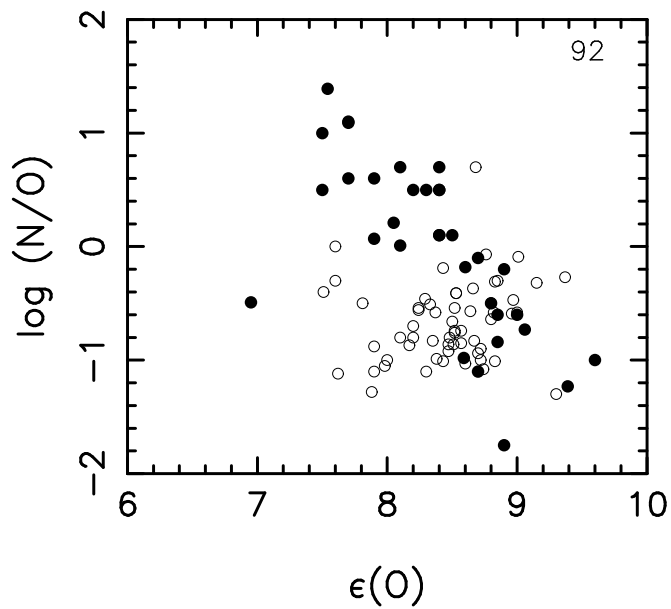}}
\caption{ The log N/O vs. $\epsilon$(O) relation.}
       \label{fig3}
\end{figure}

\subsection{Indications of dredge-up}
\begin{figure}[]
\resizebox{\hsize}{!}{\includegraphics{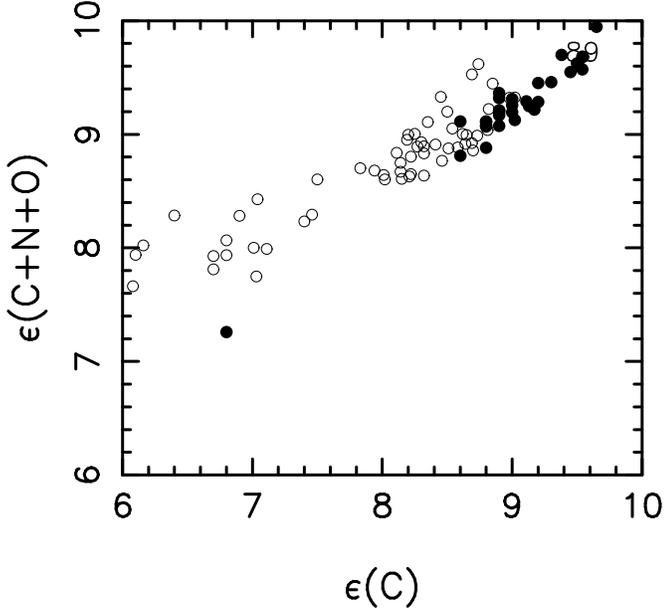}}
\caption{ $\epsilon$(C+N+O) vs. $\epsilon$(C).}
       \label{fig4} 
\end{figure}

Figure 4 shows $\epsilon$(C+N+O) vs. $\epsilon$(C). Only objects in which the abundances of the 
three elements (C, N, and O) are available are represented here. This 
plot is very similar to the plot presented by Kingsburgh \& 
Barlow\,(\cite{Kingsburgh_and_Barlow_1994}) and 
Leisy \& Dennefeld\,(\cite{Leisy_and_Dennefeld_1996}) for planetary nebulae, 
but with a larger number of points.  The objects with the highest carbon abundances, which are mainly R CrB stars and extreme helium stars, are carbon dominated. This  agrees with a scenario of C 
being produced by the triple $\alpha$  reaction and brought to the star surface 
by the third dredge-up.

\subsection{Dredge-up and stellar mass}

\begin{figure*}[]
\resizebox{\hsize}{!}{\includegraphics{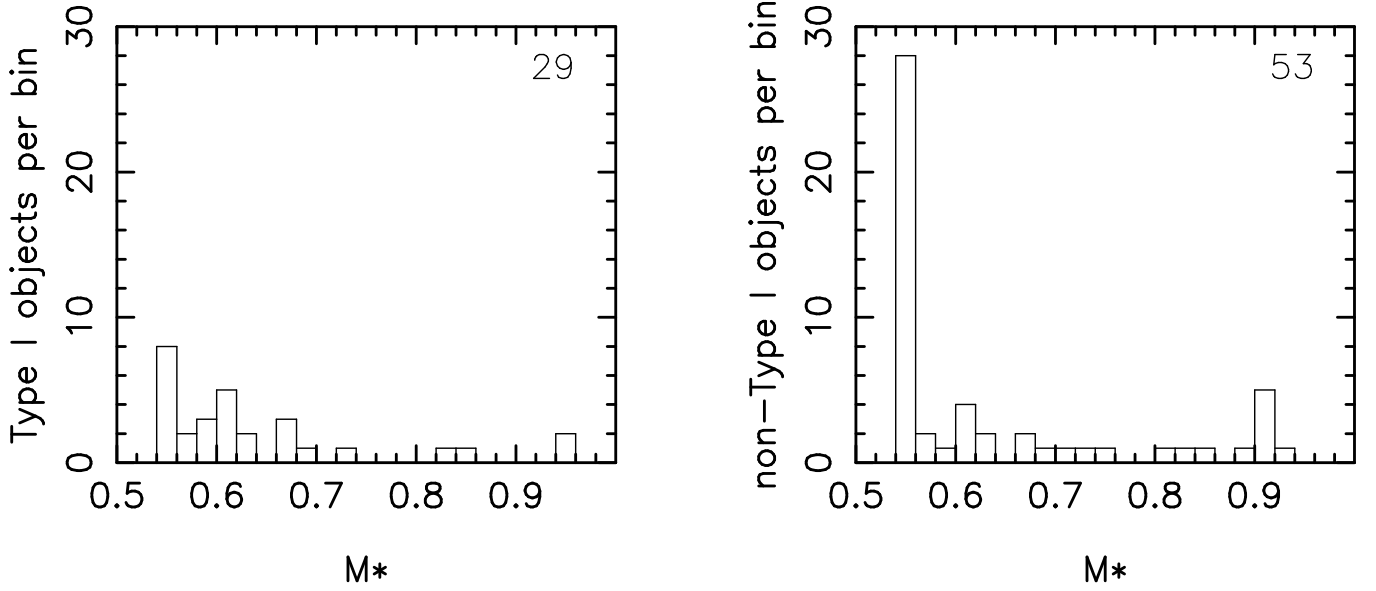}}
\caption{ Stellar mass distribution for Type I post-AGB stars (left) and 
non-Type I post-AGB stars  (right). We call Type I those objects in which N/S 
is higher than the solar (C+N)/S value.}
       \label{fig5}
\end{figure*}

A more quantitative way to define whether dredge-up mechanisms have occurred is 
to compare abundance ratios observed in post-AGB stars or planetary nebulae with  estimates of the initial abundance ratios. 
For example, Kingsburgh \& Barlow\,(\cite{Kingsburgh_and_Barlow_1994}) 
have proposed to call Type I PNe 
(understood as ``objects that have experienced envelope-burning conversion to 
nitrogen of dredged up primary carbon'') those objects in which the nitrogen 
abundance exceeds its progenitor's C+N abundance. As a proxy to the progenitor's
 C+N abundance, they use the value of C+N in the Orion nebula. Since the 
post-AGB considered here are not necessarily all close to the Sun, and because of 
abundance gradients in the Galaxy, perhaps a safer way is to compare the N/S 
value in the post-AGB stars to the solar (C+N)/S value, instead of using abundances with respect to hydrogen. For the solar 
abundances, we rely on the compilation by Lodders\,(\cite{Lodders_2003}). 
Defining as Type I those objects in which N/S is larger than the solar (C+N)/S ratio,  we show  histograms of masses of Type I (left) and non-Type 
I (right) post-AGB stars in Fig. 5. It is seen that, while Type I objects extend over the 
entire range of masses in our sample, more than half of the non-Type I objects 
have masses below 0.56\Ms. The difference in stellar mass distributions is so 
tremendous that it is highly significant, even taking into account the fact 
that error bars on stellar masses are large (as seen in Table 1  and discussed in Sect. 2.1).  The conclusion remains the same, although not so strong, if we use Zn instead of S as the metallicity indicator.  However, when considering only  R CrB and extreme helium stars, no such difference is seen.

\begin{figure}[]
\resizebox{\hsize}{!}{\includegraphics{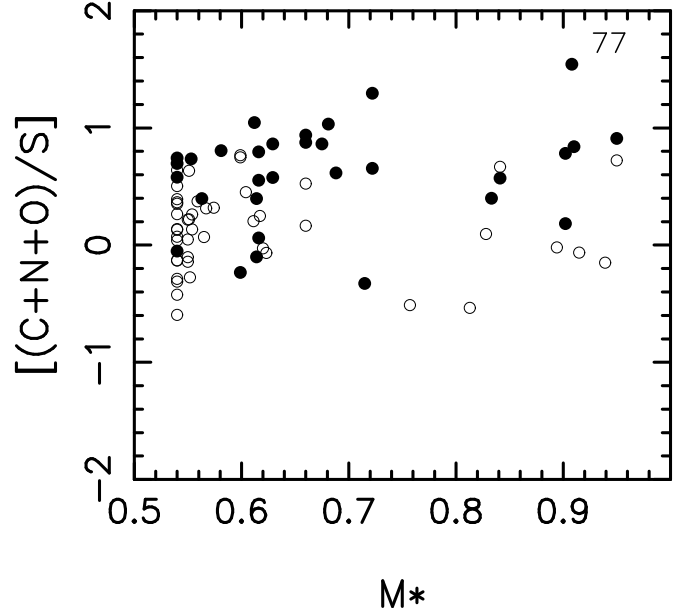}}
\caption{[(C+N+O)/S] (i.e. the logarithm of (C+N+O)/S minus the logarithm of 
the solar value of this ratio) as a function of the  mass of the post-AGB star. Objects with masses lower than  0.55\Ms\ have been placed at  0.54\Ms, objects with masses larger than  0.94\Ms\ at  0.95\Ms, }
       \label{fig6}
\end{figure}

%\begin{figure*}[]
%\resizebox{\hsize}{!}{\includegraphics{3533fig9-.ps}}
%\caption{Distribution of post-AGB stars as a function of the photospheric C/O. 
%Left: objects with a carbon-rich circumstellar chemistry. Right: objects with 
%an oxygen-rich circumstellar chemistry.}
%       \label{fig9}
%\end{figure*}

Similarly, one can identify objects that have experienced 3rd dredge-up as 
those objects in which (C+N+O)/S is larger than in the Sun. 
Figure 6 plots [(C+N+O)/S] (i.e. the logarithm of (C+N+O)/S minus the logarithm 
of the solar value of this ratio) as a function of the stellar mass. 
This diagram shows that, according to our definition ([(C+N+O)/S] $>$ 0), the 
vast majority of  post-AGB stars in our sample (about 70\%) have experienced 
3rd dredge-up. It also suggests that the mass distributions of the two 
subclasses do not differ significantly.

\subsection{Dredge-up and metallicity}

\begin{figure}[]
\resizebox{\hsize}{!}{\includegraphics{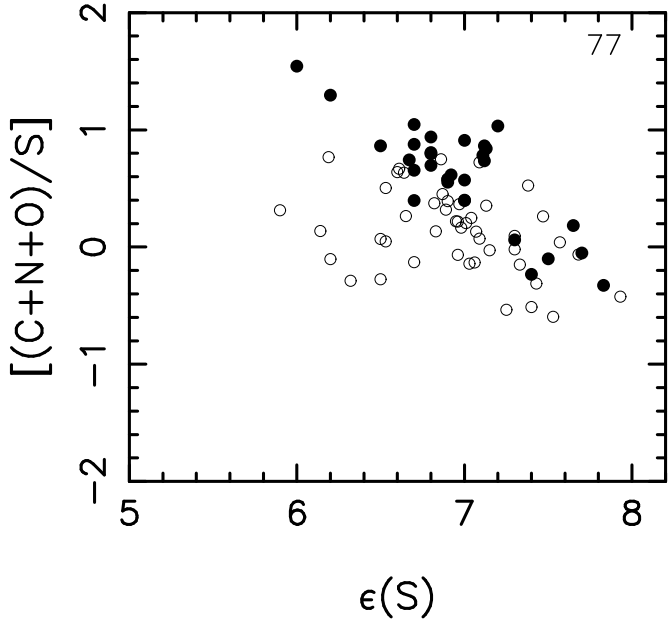}}
\caption{[(C+N+O)/S] as a function of the metallicity as measured by $\epsilon$(S).}
       \label{fig7}
\end{figure}

\begin{figure}[]
\resizebox{\hsize}{!}{\includegraphics{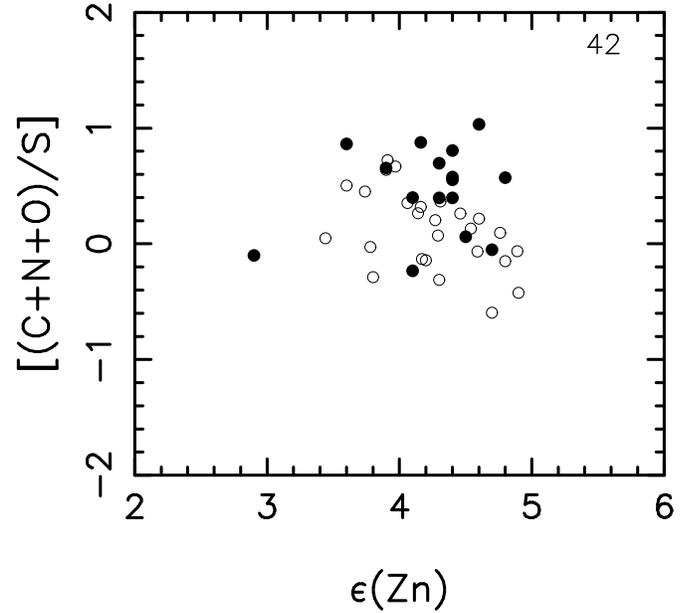}}
\caption{[(C+N+O)/S] as a function of the metallicity as measured by $\epsilon$(Zn).}
       \label{fig7}
\end{figure}

Theoretical models (e.g. Marigo 2001) predict that 3rd dredge-up is more 
important at low metallicity. Figure 7 tests this prediction by plotting 
[(C+N+O)/S] as a function of $\epsilon$(S). It shows a net decrease in the efficiency of the 3rd dredge-up as the metallicity increases, in agreement with the models. This is the first time that the observational evidence for this is so clear.  Note that, qualitatively, the same conclusion can be drawn, at least for bona-fide post-AGB stars, when using Zn instead of S as a metallicity indicator, as seen in Fig. 8, which plots [(C+N+O)/S] as a function of $\epsilon$(Zn). However, R CrB and extreme helium stars tend to have  higher [(C+N+O)/S] than bona-fide post-AGB stars of same metallicity. They also behave differently in Figs. 7 and  8, but more Zn abundance determinations would be necessary to make this clear.

\subsection{Carbon-rich versus oxygen-rich post-AGB stars.}

\begin{figure}[]
\resizebox{\hsize}{!}{\includegraphics{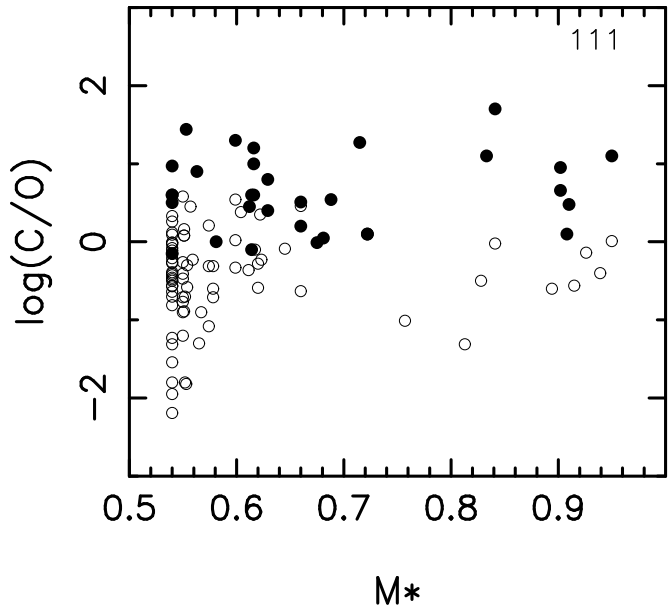}}
\caption{Photospheric C/O versus the  mass of the post-AGB stars.}
       \label{fig8}
\end{figure}

Figure 9 displays the value of C/O as a function of stellar mass. It clearly shows 
that oxygen-rich stars, at least in our sample,  tend to accumulate at the 
lowest  masses, while this is not the case for carbon-rich stars. Note also that 
the proportion of carbon-rich stars is 36\%, while the proportion of stars 
having experienced 3rd dredge-up is as large as about 70\%. Of course, these percentages should  be taken with a grain of salt since the errors in determined abundances may place an object in the wrong category. However, the difference between these percentages and the large spread of C/O, as well as [(C+N+O)/S] values, argues in favour of the number of carbon-rich stars being significantly smaller than the number of stars having experienced 3rd dredge-up. Qualitatively, this is expected, 
since 3rd dredge-up is not necessarily sufficient to produce a carbon star.

\section{Summary, open questions, and prospects}
\label{summary}

The main aim of this paper was to show the utility of  post-AGB stars to test theories of AGB nucleosynthesis. So far, the only tests of AGB nucleosynthesis based on large samples have been made using planetary nebulae. Post-AGB stars have several advantages over planetary nebulae: 1) abundances of a large variety of elements can be derived, including of $s$-process elements; 2) the abundance of carbon, an extremely important element for the diagnostics, is known with the same accuracy as the other elements, while in planetary nebulae, the carbon abundance is  significantly less reliable and more difficult to obain than that of O and N; 3) the determination of the atmospheric parameters  \Teff, and gravity $g$ allows one to estimate the mass of the post-AGB star by comparison with theoretical stellar evolutionary tracks. 

Of course, the study of post-AGB stars has its own difficulties. In particular, the abundance analysis is quite difficult  and many effects have to be considered in detail (see e.g. Asplund et al. 2000). The lack of suitable lines for reliable analysis is often a problem:  i) for cooler objects, the oxygen abundance is hard to derive, since useful
  lines of oxygen only start to show up at temperatures above  6000K;
ii) nitrogen is often derived from a few red lines, which are known to suffer
  non-LTE effects; iii) stars that are not enriched in carbon, sometimes have  only a few suitable
  carbon lines.

We  considered all those objects from the present version of the catalogue of post-AGB objects (Szczerba et al.\,\cite{Szczerba_etal_2001}, Szczerba et al., in preparation) for which photospheric chemical abundances have been determined. 
We plotted diagrams based on these abundances, similar to the ones built for planetary nebulae studies. The same trends as for planetary nebulae were found, but in a clearer fashion (e.g. N/O vs. $\epsilon$(O), revealing the effect of the ON cycle, or (C+N+O)/H vs. $\epsilon$(C) indicating the presence of objects with C produced by the triple $\alpha$ reaction). This is extremely encouraging and shows the interest of using post-AGB stars to complement planetary nebulae, despite the difficulties in abundance determinations.

Because the  post-AGB stars in our sample do not have all the same metallicities, we argued that a better indicator of third dredge-up and/or hot bottom burning is obtained by considering the photospheric abundances of C, N, O with respect to a metallicity indicator (and not with respect to H). It is therefore these ratios that we compared  with the solar ratios. We show that a convenient metallicity indicator is S (Fe cannot be used for post-AGB objects because dust depletion in former stages may have affected its present photospheric abundance).  Following the definition of Type I planetary nebulae by Kingsburgh \& Barlow (1994) but accounting for the metallicity, we define a class of Type I post-AGB stars.  We show that non-type I objects are in vast majority of low mass ($M_{\star}$ $<$ 0.56 M$_{\odot}$). We also show  clear evidence that 3rd dredge-up is more efficient at low metallicity.

We have thus demonstrated the potential of post-AGB stars to constrain the models of AGB stars and the predicted yields.  The sample of post-AGB stars is likely to grow in the near future, thanks to  ASTRO-F, which is  much more sensitive than IRAS and should allow the discovery of many infrared-excess stars among which post-AGB stars are found. This will make the use of post-AGB stars even more attractive and powerful. In the present paper, we limited ourselves to only a few elements and to simple interpretations without direct comparison to models. We did not address the question of observational biases, which should be investigated by performing simulations on models. Future studies will address other issues related to AGB nucleosynthesis, such as the production of $s$-process elements, which are more easily done with post-AGB stars than with planetary nebulae.

\begin{acknowledgements}
This work was partly supported by grant 2.P03D.017.25 of the Polish State 
Committee for Scientific Research and by the European Associated Laboratory "Astronomy Poland-France". G.S. is grateful to Beatriz Barbuy and Corinne Charbonnel for useful comments on a previous version of the paper. Comments by the referee, M. Reyniers are gratefully acknowledged.
\end{acknowledgements}

\end{document}